\documentclass[aip,jcp,reprint,noshowkeys,superscriptaddress,floatfix]{revtex4-1}
\usepackage{graphicx,dcolumn,bm,xcolor,microtype,multirow,amscd,amsmath,amssymb,amsfonts,physics,wrapfig,txfonts,siunitx}
\usepackage[version=4]{mhchem}

\newcommand{\ie}{\textit{i.e.}}

\newcommand{\alert}[1]{\textcolor{black}{#1}}
\usepackage[normalem]{ulem}

\newcommand{\mc}{\multicolumn}

\newcommand{\tabc}[1]{\multicolumn{1}{c}{#1}}

\newcommand{\Nel}{n}
\newcommand{\Norb}{N}
\newcommand{\Ndet}{N_\text{det}}

\newcommand{\hT}{\Hat{T}}
\newcommand{\hH}{\Hat{H}}
\newcommand{\hh}{\Hat{h}}

\newcommand{\hk}{\Hat{\kappa}}
\newcommand{\cre}[1]{\Hat{a}_{#1}^{\dagger}}
\newcommand{\ani}[1]{\Hat{a}_{#1}^{}}

\newcommand{\bO}{\boldsymbol{0}}
\newcommand{\bI}{\boldsymbol{1}}

\newcommand{\bH}{\boldsymbol{H}}
\newcommand{\bg}{\boldsymbol{g}}
\newcommand{\bk}{\boldsymbol{\kappa}}
\newcommand{\bc}{\boldsymbol{c}}
\newcommand{\br}{\boldsymbol{r}}

\newcommand{\cA}{\mathcal{A}}
\newcommand{\cI}{\mathcal{I}}
\newcommand{\cP}{\mathcal{P}}

\newcommand{\EHF}{E_\text{HF}}

\newcommand{\Evar}{E_\text{var}}
\newcommand{\Efinal}{E_\text{final}}
\newcommand{\Eextrap}{E_\text{extrap}}
\newcommand{\Edist}{E_\text{dist}}
\newcommand{\EPT}{E_\text{PT2}}
\newcommand{\ECIPSI}{E_\text{CIPSI}}

\newcommand{\PsiO}{\Psi_0}
\newcommand{\Psivar}{\Psi_\text{var}}
\newcommand{\MO}[1]{\phi_{#1}}

\usepackage[
	colorlinks=true,
    citecolor=blue,
    breaklinks=true
	]{hyperref}
\begin{document}

\newcommand{\LCPQ}{Laboratoire de Chimie et Physique Quantiques (UMR 5626), Universit\'e de Toulouse, CNRS, UPS, France}
\newcommand{\CEISAM}{Universit\'e de Nantes, CNRS,  CEISAM UMR 6230, F-44000 Nantes, France}

\title{Accurate full configuration interaction correlation energy estimates for five- and six-membered rings}

\author{Yann \surname{Damour}}
\affiliation{\LCPQ}
\author{Micka\"el \surname{V\'eril}}
\affiliation{\LCPQ}
\author{F\'abris \surname{Kossoski}}
\affiliation{\LCPQ}
\author{Michel \surname{Caffarel}}
\affiliation{\LCPQ}
\author{Denis \surname{Jacquemin}}
\email{Denis.Jacquemin@univ-nantes.fr}
\affiliation{\CEISAM}
\author{Anthony \surname{Scemama}}
\email{scemama@irsamc.ups-tlse.fr}
\affiliation{\LCPQ}
\author{Pierre-Fran\c{c}ois \surname{Loos}}
\email{loos@irsamc.ups-tlse.fr}
\affiliation{\LCPQ}

\begin{abstract}
Following our recent work on the benzene molecule [\href{https://doi.org/10.1063/5.0027617}{J.~Chem.~Phys.~\textbf{153}, 176101 (2020)}], itself motivated by the blind challenge of Eriksen \textit{et al.} [\href{https://doi.org/10.1021/acs.jpclett.0c02621}{J.~Phys.~Chem.~Lett.~\textbf{11}, 8922 (2020)}] on the same system, we report accurate full configuration interaction (FCI) frozen-core correlation energy estimates for twelve five- and six-membered ring molecules (cyclopentadiene, furan, imidazole, pyrrole, thiophene, benzene, pyrazine, pyridazine, pyridine, pyrimidine, s-tetrazine, and s-triazine) in the standard correlation-consistent double-$\zeta$ Dunning basis set (cc-pVDZ).
Our FCI correlation energy estimates, with estimated error smaller than 1 millihartree, are based on energetically optimized-orbital selected configuration interaction (SCI) calculations performed with the \textit{Configuration Interaction using a Perturbative Selection made Iteratively} (CIPSI) algorithm.
Having at our disposal these accurate reference energies, the respective performance and convergence properties of several popular and widely-used families of single-reference quantum chemistry methods are investigated.
In particular, we study the convergence properties of i) the M{\o}ller-Plesset perturbation series up to fifth-order (MP2, MP3, MP4, and MP5), ii) the iterative approximate coupled-cluster series CC2, CC3, and CC4, and iii) the coupled-cluster series CCSD, CCSDT, and CCSDTQ.
The performance of the ground-state gold standard CCSD(T) as well as the completely renormalized CC model, CR-CC(2,3), are also investigated.
We show that MP4 provides an interesting accuracy/cost ratio, while MP5 systematically worsen the correlation energy estimates.
In addition, CC3 outperforms CCSD(T) and CR-CC(2,3), as well as its more expensive parent CCSDT.
A similar trend is observed for the methods including quadruple excitations, where the CC4 model is shown to be slightly more accurate than CCSDTQ, both methods providing correlation energies within 2 millihartree of the FCI limit.
\end{abstract}

\maketitle

\section{Introduction}
\label{sec:intro}
Electronic structure theory relies heavily on approximations. \cite{Szabo_1996,Helgaker_2013,Jensen_2017}  
Loosely speaking, to make any method practical, three main approximations are typically enforced.
The first fundamental approximation, known as the Born-Oppenheimer (or clamped-nuclei) approximation, consists in assuming that the motion of nuclei and electrons are decoupled. \cite{Born_1927}
The nuclei coordinates can then be treated as parameters in the electronic Hamiltonian.
The second central approximation which makes calculations computationally achievable is the basis set approximation where one introduces a set of pre-defined basis functions to represent the many-electron wave function of the system.
In most molecular calculations, a set of one-electron, atom-centered Gaussian basis functions are introduced to expand the so-called one-electron molecular orbitals which are then used to build the many-electron Slater determinant(s).
The third and most relevant approximation in the present context is the ansatz (or form) of the electronic wave function $\Psi$.
For example, in configuration interaction (CI) methods, the wave function is expanded as a linear combination of Slater determinants, while in (single-reference) coupled-cluster (CC) theory, \cite{Cizek_1966,Paldus_1972,Crawford_2000,Piecuch_2002b,Bartlett_2007,Shavitt_2009} a reference Slater determinant $\PsiO$ [usually taken as the Hartree-Fock (HF) wave function] is multiplied by a wave operator defined as the exponentiated excitation operator $\hT = \sum_{k=1}^\Nel \hT_k$ (where $\Nel$ is the number of electrons and $\hT_k$ is $k$th-degree excitation operator).

\begin{figure*}
	\includegraphics[width=0.8\linewidth]{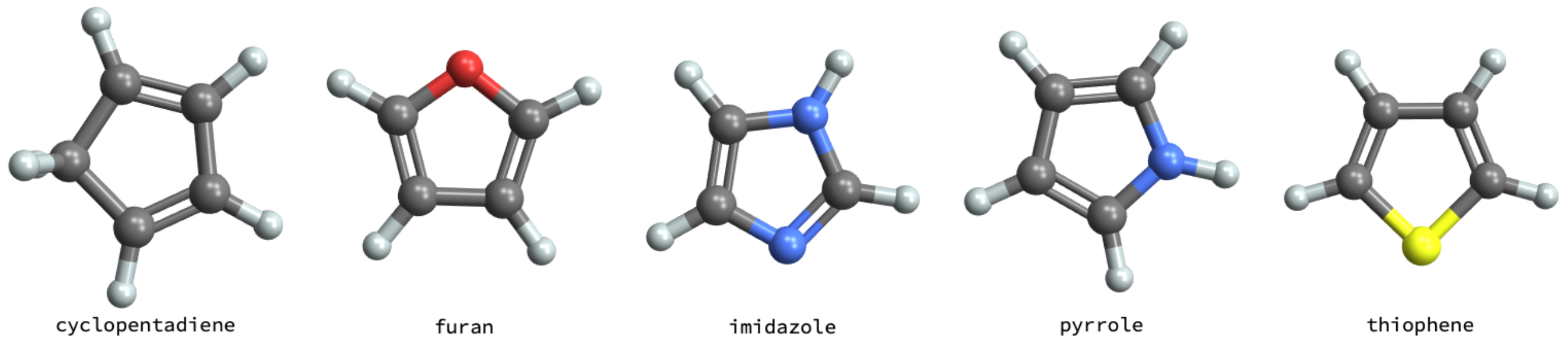}
	\includegraphics[width=\linewidth]{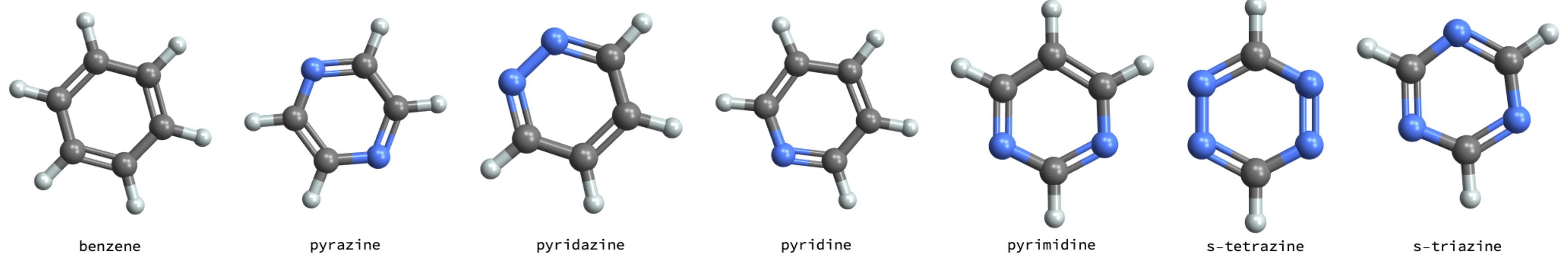}
	\caption{
	Five-membered rings (top) and six-membered rings (bottom) considered in this study.
	\label{fig:mol}}
\end{figure*}

The truncation of $\hT$ allows to define a hierarchy of non-variational and size-extensive methods with increasing levels of accuracy:
CC with singles and doubles (CCSD), \cite{Cizek_1966,Purvis_1982} CC with singles, doubles, and triples (CCSDT), \cite{Noga_1987a,Scuseria_1988} CC with singles, doubles, triples, and quadruples (CCSDTQ), \cite{Oliphant_1991,Kucharski_1992} with corresponding formal computational scalings of $\order*{\Norb^{6}}$, $\order*{\Norb^{8}}$, and $\order*{\Norb^{10}}$, respectively (where $\Norb$ denotes the number of orbitals).
Parallel to the ``complete'' CC series presented above, an alternative family of approximate iterative CC models has been developed by the Aarhus group in the context of CC response theory \cite{Christiansen_1998} where one skips the most expensive terms and avoids the storage of the higher-excitation amplitudes: CC2, \cite{Christiansen_1995a} CC3, \cite{Christiansen_1995b,Koch_1997} and CC4. \cite{Kallay_2005,Matthews_2021}
These iterative methods scale as $\order*{\Norb^{5}}$, $\order*{\Norb^{7}}$, and $\order*{\Norb^{9}}$, respectively, and can be seen as cheaper approximations of CCSD, CCSDT, and CCSDTQ.
Coupled-cluster methods have been particularly successful at computing accurately ground- and excited-state properties for small- and medium-sized molecules.
\cite{Kallay_2003,Kallay_2004a,Gauss_2006,Kallay_2006,Gauss_2009,Chrayteh_2021,Sarkar_2021}

A similar systematic truncation strategy can be applied to CI methods leading to the well-established family of methods known as CISD, CISDT, CISDTQ, \ldots~where one systematically increases the maximum excitation degree of the determinants taken into account.
Except for full CI (FCI) where all determinants from the Hilbert space (\ie, with excitation degree up to $\Nel$) are considered, truncated CI methods are variational but lack size-consistency.
The non-variationality of truncated CC methods being, in practice, less of an issue than the size-inconsistency of the truncated CI methods, the formers have naturally overshadowed the latters in the electronic structure landscape.
However, a different strategy recently came back in the limelight in the context of CI methods. \cite{Bender_1969,Whitten_1969,Huron_1973,Shih_1978,Buenker_1978,Evangelisti_1983,Cimiraglia_1985,Cimiraglia_1987,Illas_1988,Povill_1992,Abrams_2005,Bunge_2006}
Indeed, selected CI (SCI) methods, \cite{Booth_2009,Giner_2013,Evangelista_2014,Giner_2015,Caffarel_2016b,Holmes_2016,Tubman_2016,Liu_2016,Ohtsuka_2017,Zimmerman_2017,Coe_2018,Garniron_2018} where one iteratively selects the important determinants from the FCI space (usually) based on a perturbative criterion, has been recently shown to be highly successful in order to produce reference energies for both ground and excited states in small- and medium-sized molecules \cite{Caffarel_2014,Caffarel_2016a,Scemama_2016,Holmes_2017,Li_2018,Scemama_2018,Scemama_2018b,Li_2020,Loos_2018a,Chien_2018,Loos_2019,Loos_2020b,Loos_2020c,Loos_2020e,Garniron_2019,Eriksen_2020,Yao_2020,Williams_2020,Veril_2021,Loos_2021} thanks to efficient deterministic, stochastic, or hybrid algorithms well suited for massive parallelization.
We refer the interested reader to Refs.~\onlinecite{Loos_2020a,Eriksen_2021} for recent reviews.
SCI methods are based on a well-known fact: amongst the very large number of determinants contained in the FCI space, only a tiny fraction of them significantly contributes to the energy (see, for example, Ref.~\onlinecite{Ivanic_2001}).
Accordingly, the SCI+PT2 family of methods performs a sparse exploration of the FCI space by selecting iteratively only the most energetically relevant determinants of the variational space and supplementing it with a second-order perturbative correction (PT2). \cite{Huron_1973,Garniron_2017,Sharma_2017,Garniron_2018,Garniron_2019}
Although the formal scaling of such algorithms remains exponential, the prefactor is greatly reduced which explains their current attractiveness in the electronic structure community thanks to their much wider applicability than their standard FCI parent.
Note that, very recently, several groups \cite{Aroeira_2021,Lee_2021,Magoulas_2021} have coupled CC and SCI methods via the externally-corrected CC methodology, \cite{Paldus_2017} showing promising performances for weakly and strongly correlated systems.

A rather different strategy in order to reach the holy grail FCI limit is to resort to M{\o}ller-Plesset (MP) perturbation theory, \cite{Moller_1934} whose popularity originates from its black-box nature, size-extensivity, and relatively low computational requirement, making it easily applied to a broad range of molecular systems. 
Again, at least in theory, one can obtain the exact energy of the system by ramping up the degree of the perturbative series. \cite{Marie_2021a}
The second-order M{\o}ller-Plesset (MP2) method \cite{Moller_1934} [which scales as $\order*{\Norb^{5}}$] has been broadly adopted in quantum chemistry for several decades, and is now included in the increasingly popular double-hybrid functionals \cite{Grimme_2006} alongside exact exchange. 
Its higher-order variants [MP3, \cite{Pople_1976} MP4, \cite{Krishnan_1980} MP5, \cite{Kucharski_1989} and MP6 \cite{He_1996a,He_1996b} which scale as $\order*{\Norb^{6}}$, $\order*{\Norb^{7}}$, $\order*{\Norb^{8}}$, and $\order*{\Norb^{9}}$ respectively] have been investigated much more scarcely.
However, it is now widely recognized that the series of MP approximations might show erratic, slowly convergent, or divergent behavior that limits its applicability and systematic improvability. \cite{Laidig_1985,Knowles_1985,Handy_1985,Gill_1986,Laidig_1987,Nobes_1987,Gill_1988,Gill_1988a,Lepetit_1988,Malrieu_2003,Marie_2021a}
Again, MP perturbation theory and CC methods can be coupled.
The most iconic example of such coupling, namely the CCSD(T) method, \cite{Raghavachari_1989} includes iteratively the single and double excitations and perturbatively (from MP4 and partially MP5) the triple excitations, leading to the so-called ``gold-standard'' of quantum chemistry for weakly correlated systems thanks to its excellent accuracy/cost ratio.

Motivated by the recent blind test of Eriksen \textit{et al.}\cite{Eriksen_2020}~reporting the performance of a large panel of emerging electronic structure methods [the many-body expansion FCI (MBE-FCI), \cite{Eriksen_2017,Eriksen_2018,Eriksen_2019a,Eriksen_2019b} adaptive sampling CI (ASCI), \cite{Tubman_2016,Tubman_2018,Tubman_2020} iterative CI (iCI), \cite{Liu_2014,Liu_2016,Lei_2017,Zhang_2020} semistochastic heat-bath CI (SHCI), \cite{Holmes_2016,Holmes_2017,Sharma_2017} the full coupled-cluster reduction (FCCR), \cite{Xu_2018,Xu_2020} density-matrix renormalization group (DMRG), \cite{White_1992,White_1993,Chan_2011} adaptive-shift FCI quantum Monte Carlo (AS-FCIQMC), \cite{Booth_2009,Cleland_2010,Ghanem_2019} and cluster-analysis-driven FCIQMC (CAD-FCIQMC) \cite{Deustua_2017,Deustua_2018}] on the non-relativistic frozen-core correlation energy of the benzene molecule in the standard correlation-consistent double-$\zeta$ Dunning basis set (cc-pVDZ), some of us have recently investigated the performance of the SCI method known as \textit{Configuration Interaction using a Perturbative Selection made Iteratively} (CIPSI). \cite{Huron_1973,Giner_2013,Giner_2015,Garniron_2018,Garniron_2019} on the very same system \cite{Loos_2020e} [see also Ref.~\onlinecite{Lee_2020} for a study of the performance of phaseless auxiliary-field quantum Monte Carlo (ph-AFQMC) \cite{Motta_2018}].
In the continuity of this recent work, we report here a large extension by accurately estimating the (frozen-core) FCI/cc-pVDZ correlation energy of twelve cyclic molecules (cyclopentadiene, furan, imidazole, pyrrole, thiophene, benzene, pyrazine, pyridazine, pyridine, pyrimidine, s-tetrazine, and s-triazine) with the help of CIPSI employing energetically-optimized orbitals at the same level of theory. \cite{Yao_2020,Yao_2021}
These systems are depicted in Fig.~\ref{fig:mol}.
This set of molecular systems corresponds to Hilbert spaces with sizes ranging from $10^{29}$ to $10^{36}$.
In addition to CIPSI, the performance and convergence properties of several series of methods are investigated.
In particular, we study i) the MP perturbation series up to fifth-order (MP2, MP3, MP4, and MP5), ii) the CC2, CC3, and CC4 approximate series, and ii) the ``complete'' CC series up to quadruples (\ie, CCSD, CCSDT, and CCSDTQ).
The performance of the ground-state gold standard CCSD(T) as well as the completely renormalized (CR) CC model, CR-CC(2,3), \cite{Kowalski_2000a,Kowalski_2000b,Piecuch_2002a,Piecuch_2002b,Piecuch_2005} are also investigated.
\alert{From a theoretical point of view, one would expect the following ranking: MP2 $<$ CC2 $<$ MP3 $<$ CCSD $<$ MP4 $<$ CCSD(T) $<$ CR-CC(2,3) $<$ CC3 $<$ CCSDT $<$ MP5 $<$ CC4 $<$ CCSDTQ. But, as we shall see below, this ranking is slightly altered for the present systems.}

The present manuscript is organized as follows.
In Sec.~\ref{sec:OO-CIPSI}, we provide theoretical details about the CIPSI algorithm and the orbital optimization procedure employed here.
Section \ref{sec:compdet} deals with computational details concerning geometries, basis sets, and methods.
In Sec.~\ref{sec:res}, we report our reference FCI correlation energies for the five-membered and six-membered cyclic molecules obtained thanks to extrapolated orbital-optimized CIPSI calculations (Sec.~\ref{sec:cipsi_res}).
These reference correlation energies are then used to benchmark and study the convergence properties of various perturbative and CC methods (Sec.~\ref{sec:mpcc_res}).
Finally, we draw our conclusions in Sec.~\ref{sec:ccl}.

\section{CIPSI with optimized orbitals}
\label{sec:OO-CIPSI}

Here, we provide key details about the CIPSI method \cite{Huron_1973,Garniron_2019} as well as the orbital optimization procedure which has been shown to be highly effective in the context of SHCI by Umrigar and coworkers. \cite{Eriksen_2020,Yao_2020,Yao_2021}
Although we focus on the ground state, the present discussion can be easily extended to excited states. \cite{Scemama_2019,Veril_2021}

At the $k$th iteration, the total CIPSI energy $\ECIPSI^{(k)}$ is defined as the sum of the variational energy
\begin{equation}
	\Evar^{(k)} = \frac{\mel*{\Psivar^{(k)}}{\hH}{\Psivar^{(k)}}}{\braket*{\Psivar^{(k)}}{\Psivar^{(k)}}}
\end{equation}
and a second-order perturbative energy correction
\begin{equation}
	\EPT^{(k)}
	= \sum_{\alpha \in \cA_k} e_{\alpha}^{(k)}
	= \sum_{\alpha \in \cA_k} \frac{\abs*{\mel*{\Psivar^{(k)}}{\hH}{\alpha}}^2}{\Evar^{(k)} - \mel*{\alpha}{\hH}{\alpha}},
\end{equation}
where $\hH$ is the (non-relativistic) electronic Hamiltonian,
\begin{equation}
\label{eq:Psivar}
	\Psivar^{(k)} = \sum_{I \in \cI_k} c_I^{(k)} \ket*{I}
\end{equation}
is the variational wave function, $\cI_k$ is the set of internal determinants $\ket*{I}$ and $\cA_k$ is the set of external determinants (or perturbers) $\ket*{\alpha}$ which do not belong to the variational space at the $k$th iteration but are linked to it via a nonzero matrix element, \ie, $\mel*{\Psivar^{(k)}}{\hH}{\alpha} \neq 0$.
The sets $\cI_k$ and $\cA_k$ define, at the $k$th iteration, the internal and external spaces, respectively.
In the selection step, the perturbers corresponding to the largest $\abs*{e_{\alpha}^{(k)}}$ values are then added to the variational space at the next iteration.
In our implementation, the size of the variational space is roughly doubled at each iteration.
Hereafter, we label these iterations over $k$ which consist in enlarging the variational space as \textit{macroiterations}.
In practice, $\Evar^{(k)}$ is the lowest eigenvalue of the $\Ndet^{(k)} \times \Ndet^{(k)}$ CI matrix with elements $\mel{I}{\hH}{J}$ obtained via Davidson's algorithm. \cite{Davidson_1975}
The magnitude of $\EPT^{(k)}$ provides, at iteration $k$, a qualitative idea of the distance to the FCI limit. \cite{Garniron_2018}
We then linearly extrapolate, using large variational wave functions, the CIPSI energy to $\EPT = 0$ (which effectively corresponds to the FCI limit).
Further details concerning the extrapolation procedure are provided below (see Sec.~\ref{sec:res}).

Orbital optimization techniques at the SCI level are theoretically straightforward, but practically challenging.
Some of the technology presented here has been borrowed from complete-active-space self-consistent-field (CASSCF) methods \cite{Werner_1980,Werner_1985,Sun_2017,Kreplin_2019,Kreplin_2020} but one of the strength of SCI methods is that one does not need to select an active space and to classify orbitals as active, inactive, and virtual orbitals.
Here, we detail our orbital optimization procedure within the CIPSI algorithm and we assume that the variational wave function is normalized, \ie, $\braket*{\Psivar}{\Psivar} = 1$.

As stated in Sec.~\ref{sec:intro}, $\Evar$ depends on both the CI coefficients $\{ c_I \}_{1 \le I \le \Ndet}$ [see Eq.~\eqref{eq:Psivar}] but also on the orbital rotation parameters $\{\kappa_{pq}\}_{1 \le p,q \le \Norb}$.
Motivated by cost saving arguments, we have chosen to optimize separately the CI and orbital coefficients by alternatively diagonalizing the CI matrix after each selection step and then rotating the orbitals until the variational energy, for a given number of determinants, is minimal. 
We refer the interested reader to the recent work of Yao and Umrigar for a detailed comparison of coupled, uncoupled, and partially-coupled optimizations within SCI methods. \cite{Yao_2021}
Following the standard procedure, \cite{Helgaker_2013} we conveniently rewrite the variational energy as
\begin{equation}
\label{eq:Evar_c_k}
	\Evar(\bc,\bk) = \mel{\Psivar}{e^{\hk} \hH e^{-\hk}}{\Psivar},
\end{equation}
where $\bc$ gathers the CI coefficients, $\bk$ the orbital rotation parameters, and
\begin{equation}
	\hk = \sum_{p < q} \sum_{\sigma} \kappa_{pq} \qty(\cre{p\sigma} \ani{q\sigma} - \cre{q\sigma} \ani{p\sigma})
\end{equation}
is a real-valued one-electron antisymmetric operator, which creates an orthogonal transformation of the orbital coefficients when exponentiated, $\ani{p\sigma}$ ($\cre{p\sigma}$) being the second quantization annihilation (creation) operator which annihilates (creates) a spin-$\sigma$ electron in the real-valued spatial orbital $\MO{p}(\br)$. \cite{Helgaker_2013}

Applying the Newton-Raphson method by Taylor-expanding the variational energy to second order around $\bk = \bO$, \ie,
\begin{equation}
	\label{eq:EvarTaylor}
	\Evar(\bc,\bk) \approx \Evar(\bc,\bO) + \bg \cdot \bk + \frac{1}{2} \bk^{\dag} \cdot \bH \cdot \bk,
\end{equation}
one can iteratively minimize the variational energy with respect to the parameters $\kappa_{pq}$ by setting
\begin{equation}
	\label{eq:kappa_newton}
	\bk = - \bH^{-1} \cdot \bg,
\end{equation}
where $\bg$ and $\bH$ are the orbital gradient and Hessian matrices, respectively, both evaluated at $\bk = \bO$.
Their elements are explicitly given by the following expressions: \cite{Bozkaya_2011,Henderson_2014a}
\begin{equation}
\begin{split}
	g_{pq}
	&=  \left. \pdv{\Evar(\bc,\bk)}{\kappa_{pq}}\right|_{\bk=\bO}
	\\
	&= \sum_{\sigma} \mel{\Psivar}{\comm*{\cre{p\sigma} \ani{q\sigma} - \cre{q\sigma} \ani{p\sigma}}{\hH}}{\Psivar}
	\\
	&= \cP_{pq} \qty[ \sum_r \left( h_p^r \ \gamma_r^q - h_r^q \ \gamma_p^r \right) + \sum_{rst} \qty( v_{pt}^{rs} \Gamma_{rs}^{qt} - v_{rs}^{qt} \Gamma_{pt}^{rs} ) ],
\end{split}
\end{equation}
and
\begin{equation}
\begin{split}
	H_{pq,rs}
	& = \left. \pdv{\Evar(\bc,\bk)}{\kappa_{pq}}{\kappa_{rs}}\right|_{\bk=\bO}
	\\
	& = \cP_{pq} \cP_{rs} \Bigg\{
	\frac{1}{2} \sum_{\sigma\sigma'} \mel*{\Psivar}{\comm*{\cre{r \sigma'} \ani{s \sigma'}}{\comm*{\cre{p \sigma} \ani{q \sigma}}{\hH}}}{\Psivar}
	\\
	& \phantom{\cP_{pq} \cP_{rs} \Bigg\{} + \frac{1}{2} \sum_{\sigma\sigma'} \mel*{\Psivar}{\comm*{\cre{p \sigma} \ani{q \sigma}}{\comm*{\cre{r \sigma'} \ani{s \sigma'}}{\hH}}}{\Psivar}
	\Bigg\}
	\\
	& = \cP_{pq} \cP_{rs} \Bigg\{
	\frac{1}{2} \sum_u \qty[ \delta_{qr}(h_p^u \gamma_u^s + h_u^s \gamma_p^u) + \delta_{ps}(h_r^u \gamma_u^q + h_u^q \gamma_u^r)]
	\\
	& \phantom{\cP_{pq} \cP_{rs} \Bigg\{} - (h_p^s \gamma_r^q + h_r^q \gamma_p^s)
	\\
	& \phantom{\cP_{pq} \cP_{rs} \Bigg\{} + \frac{1}{2} \sum_{tuv} \delta_{qr}(v_{pt}^{uv} \Gamma_{uv}^{st} + v_{uv}^{st} \Gamma_{pt}^{uv})
	\\
	& \phantom{\cP_{pq} \cP_{rs} \Bigg\{} + \frac{1}{2} \sum_{tuv} \delta_{ps}(v_{uv}^{qt} \Gamma_{rt}^{uv} + v_{rt}^{uv}\Gamma_{uv}^{qt})]
	\\
	& \phantom{\cP_{pq} \cP_{rs} \Bigg\{} + \sum_{uv} (v_{pr}^{uv} \Gamma_{uv}^{qs} + v_{uv}^{qs}  \Gamma_{ps}^{uv})
	\\
	& \phantom{\cP_{pq} \cP_{rs} \Bigg\{} - \sum_{tu} (v_{pu}^{st} \Gamma_{rt}^{qu}+v_{pu}^{tr} \Gamma_{tr}^{qu}+v_{rt}^{qu}\Gamma_{pu}^{st} + v_{tr}^{qu}\Gamma_{pu}^{ts})]
	\Bigg\},
\end{split}
\end{equation}
where $\delta_{pq}$ is the Kronecker delta, $\cP_{pq} = 1 - (p \leftrightarrow q)$ is a permutation operator,
\begin{subequations}
\begin{gather}
	\label{eq:one_dm}
	\gamma_p^q = \sum_{\sigma} \mel{\Psivar}{\hat{a}_{p \sigma}^{\dagger} \hat{a}_{q \sigma}^{}}{\Psivar},
	\\
	\label{eq:two_dm}
	\Gamma_{pq}^{rs} = \sum_{\sigma \sigma'} \mel{\Psivar}{\cre{p\sigma} \cre{r\sigma'} \ani{s\sigma'} \ani{q\sigma}}{\Psivar}
\end{gather}
\end{subequations}
are the elements of the one- and two-electron density matrices, and
\begin{subequations}
\begin{gather}
	\label{eq:one}
	h_p^q = \int \MO{p}(\br) \, \hh(\br) \, \MO{q}(\br) d\br,
	\\
	\label{eq:two}
	v_{pq}^{rs} = \iint  \MO{p}(\br_1) \MO{q}(\br_2) \frac{1}{\abs*{\br_1 - \br_2}} \MO{r}(\br_1) \MO{s}(\br_2) d\br_1 d\br_2
\end{gather}
\end{subequations}
are the one- and two-electron integrals, respectively.

Because the size of the CI space is much larger than the orbital space, for each macroiteration, we perform multiple \textit{microiterations} which consist in iteratively minimizing the variational energy \eqref{eq:Evar_c_k} with respect to the $\Norb(\Norb-1)/2$ independent orbital rotation parameters for a fixed set of determinants.
After each microiteration (\ie, orbital rotation), the one- and two-electron integrals [see Eqs.~\eqref{eq:one} and \eqref{eq:two}] have to be updated. 
Moreover, the CI matrix must be re-diagonalized and new one- and two-electron density matrices [see Eqs.~\eqref{eq:one_dm} and \eqref{eq:two_dm}] have to be computed.
Microiterations are stopped when a stationary point is found, \ie, $\norm{\bg}_\infty < \tau$, where $\tau$ is a user-defined threshold which has been set to $10^{-4}$ a.u.~in the present study, and a new CIPSI selection step is performed.
Note that a tight convergence is not critical here as a new set of microiterations is performed at each macroiteration and a new production CIPSI run is performed from scratch using the final set of orbitals (see Sec.~\ref{sec:compdet}).
This procedure might sound computationally expensive but one has to realize that the microiterations are usually performed only for relatively compact variational spaces.
Therefore, the computational bottleneck of this approach remains the diagonalization of the CI matrix for very large variational spaces.

To enhance the convergence of the microiteration process, we employ an adaptation of the Newton-Raphson method known as ``trust region''. \cite{Nocedal_1999}
This popular variant defines a region where the quadratic approximation \eqref{eq:EvarTaylor} is an adequate representation of the objective energy function \eqref{eq:Evar_c_k} and it evolves during the optimization process in order to preserve the adequacy via a constraint on the step size preventing it from overstepping, \ie, $\norm{\bk} \leq \Delta$, where $\Delta$ is the trust radius.
By introducing a Lagrange multiplier $\lambda$ to control the trust-region size, one replaces Eq.~\eqref{eq:kappa_newton} by $\bk = - (\bH + \lambda \bI)^{-1} \cdot \bg$.
The addition of the level shift $\lambda \geq 0$ removes the negative eigenvalues and ensures the positive definiteness of the Hessian matrix by reducing the step size.
By choosing the right value of $\lambda$, $\norm{\bk}$ is constrained within a hypersphere of radius $\Delta$ and is able to evolve from the Newton direction at $\lambda = 0$ to the steepest descent direction as $\lambda$ grows.
The evolution of the trust radius during the optimization and the use of a condition to reject the step when the energy rises ensure the convergence of the algorithm.
More details can be found in Ref.~\onlinecite{Nocedal_1999}.

\begin{figure*}
	\includegraphics[width=0.24\textwidth]{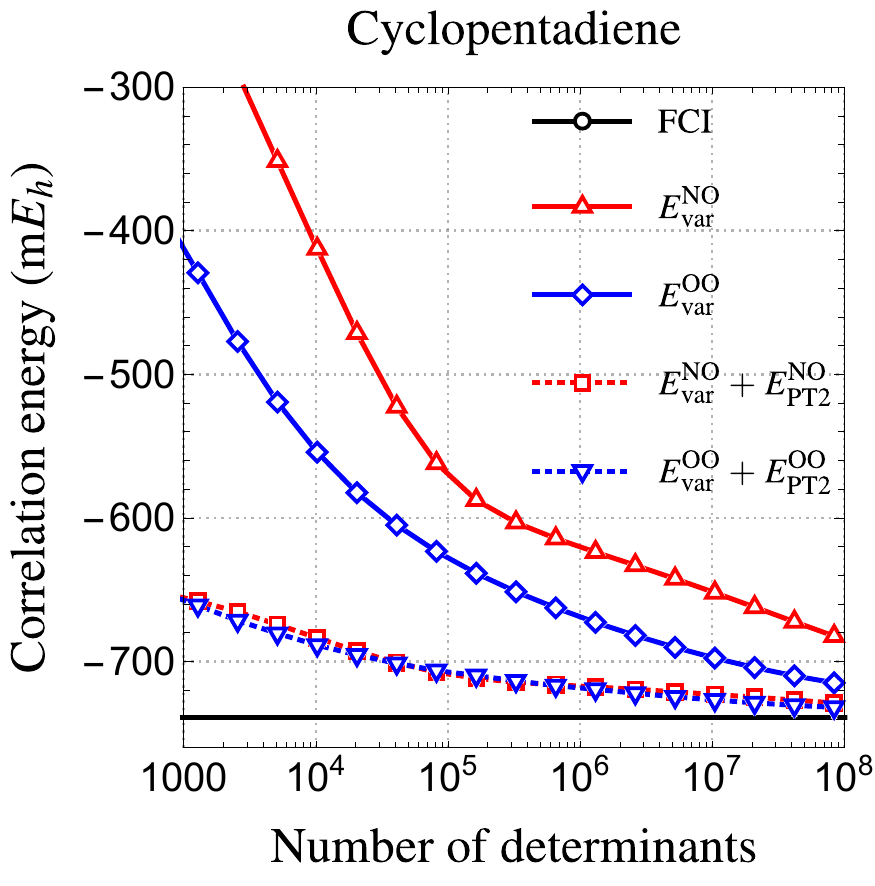}
	\includegraphics[width=0.24\textwidth]{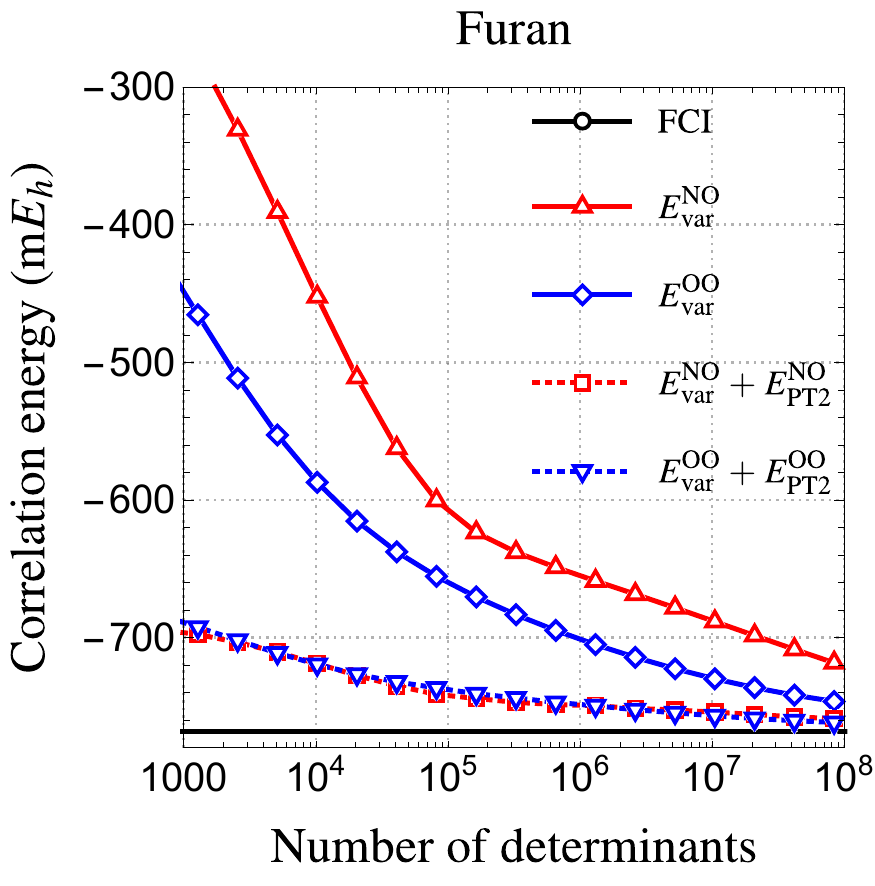}
	\includegraphics[width=0.24\textwidth]{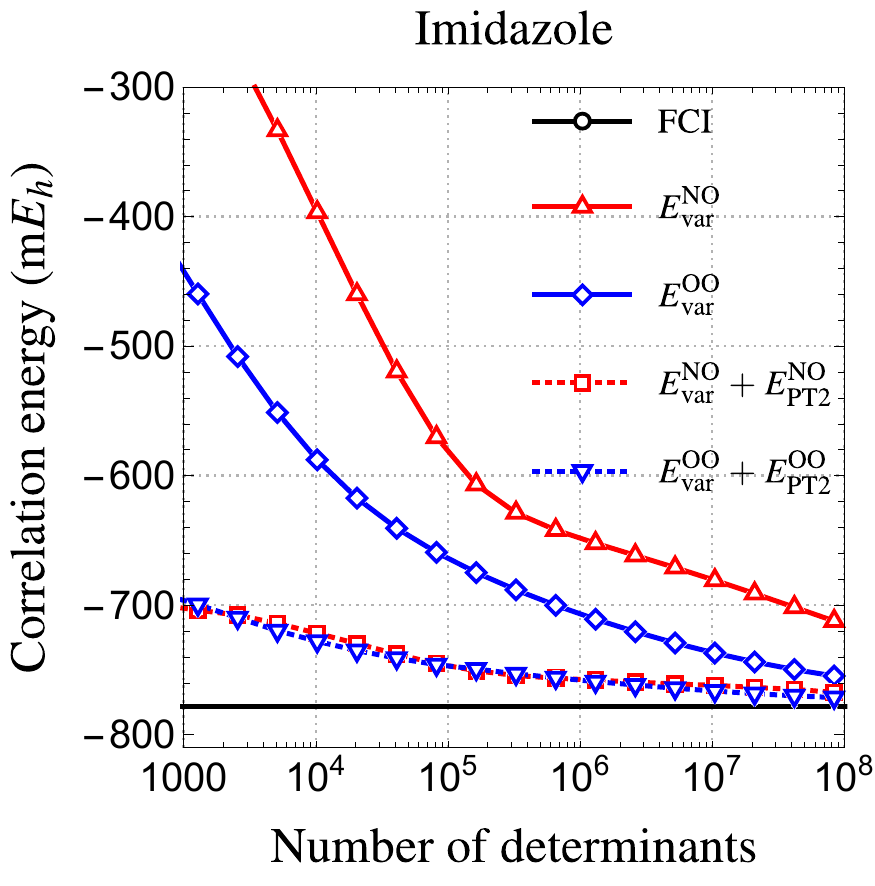}
	\includegraphics[width=0.24\textwidth]{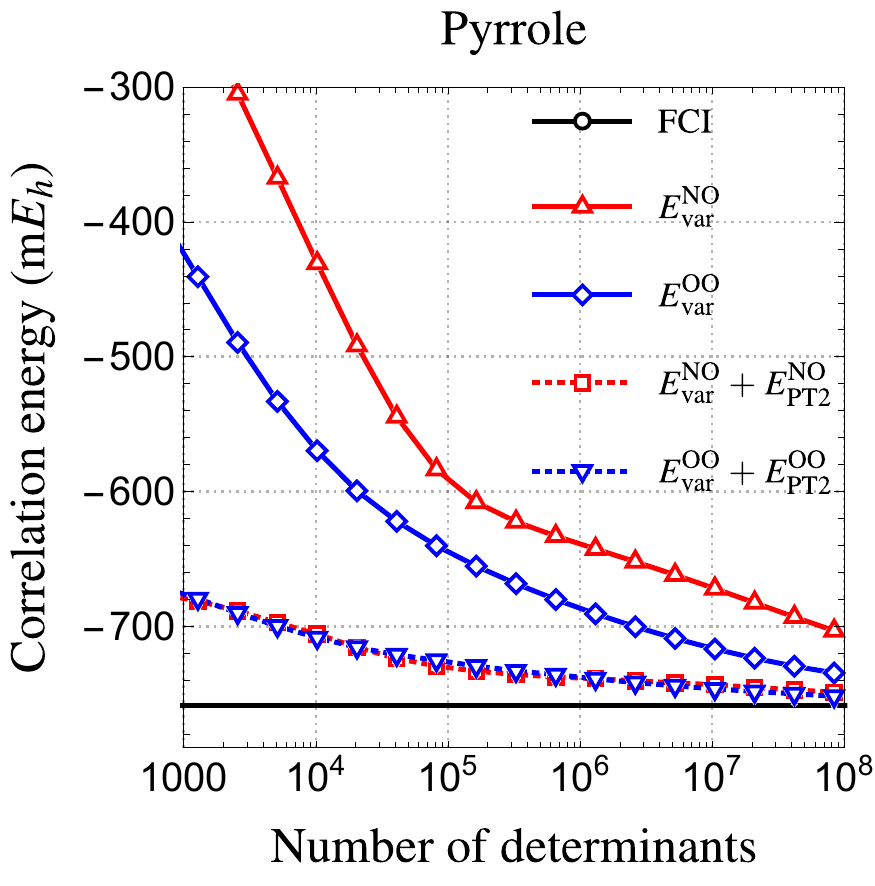}
	\\
	\includegraphics[width=0.24\textwidth]{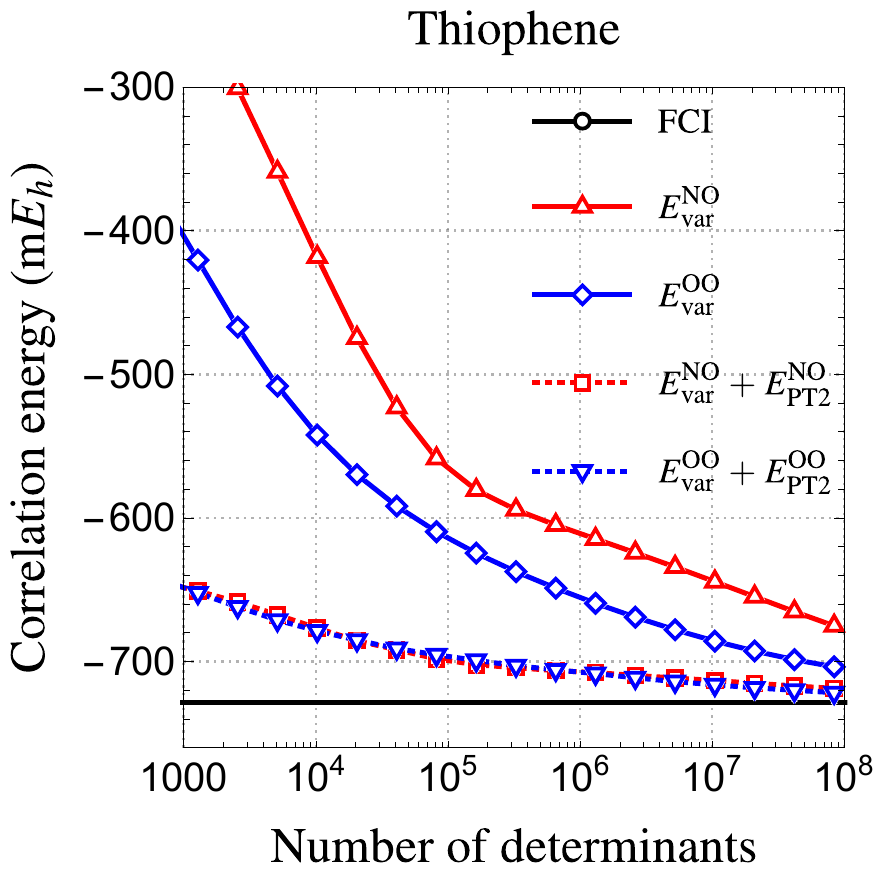}
	\includegraphics[width=0.24\textwidth]{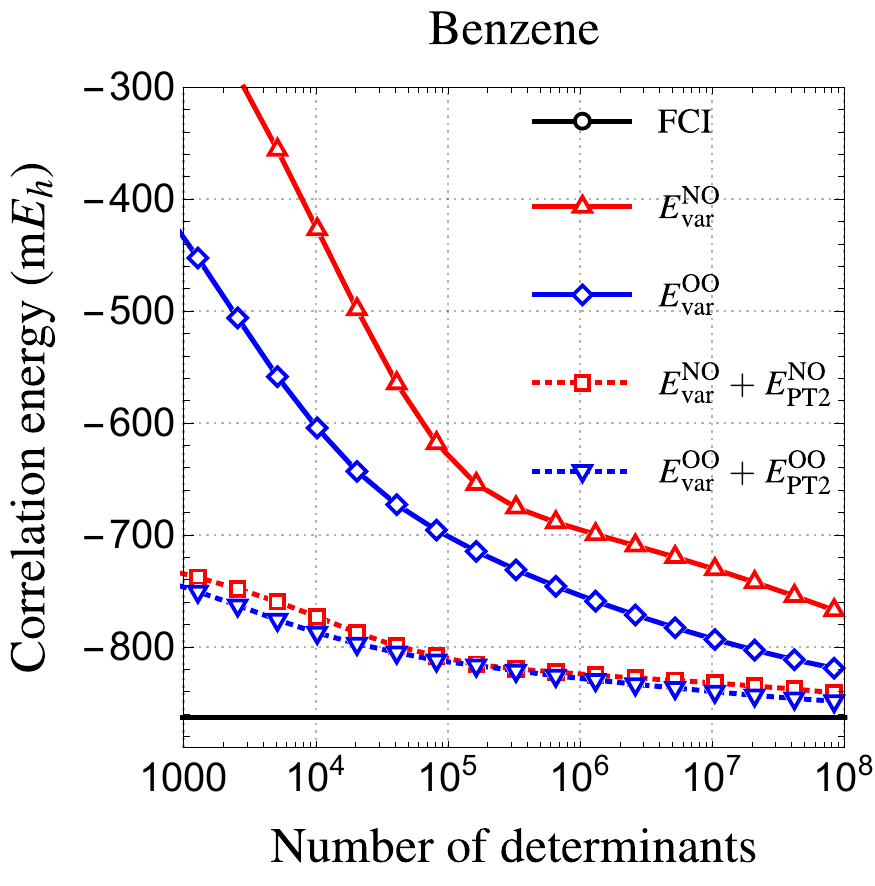}
	\includegraphics[width=0.24\textwidth]{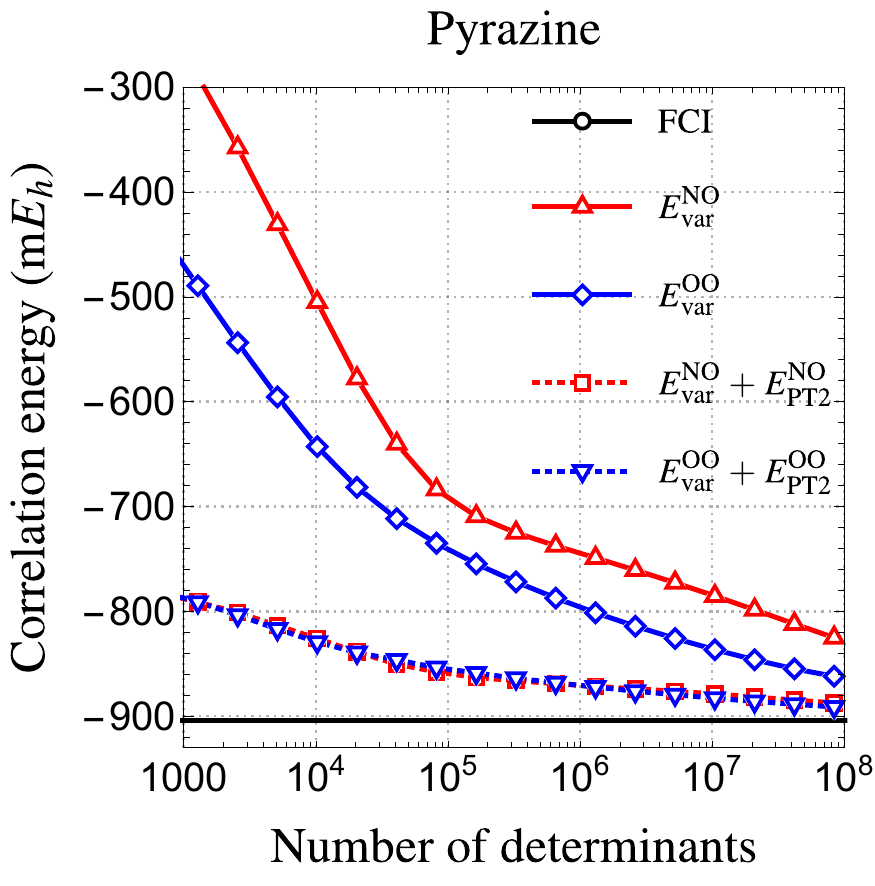}
	\includegraphics[width=0.24\textwidth]{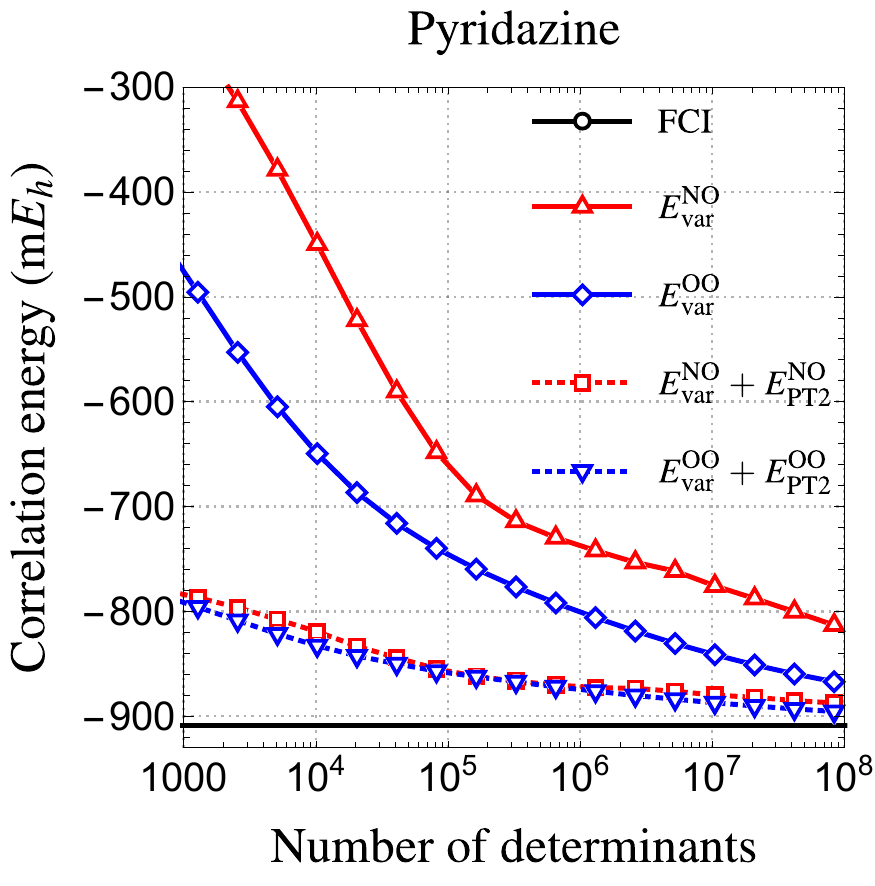}
	\\
	\includegraphics[width=0.24\textwidth]{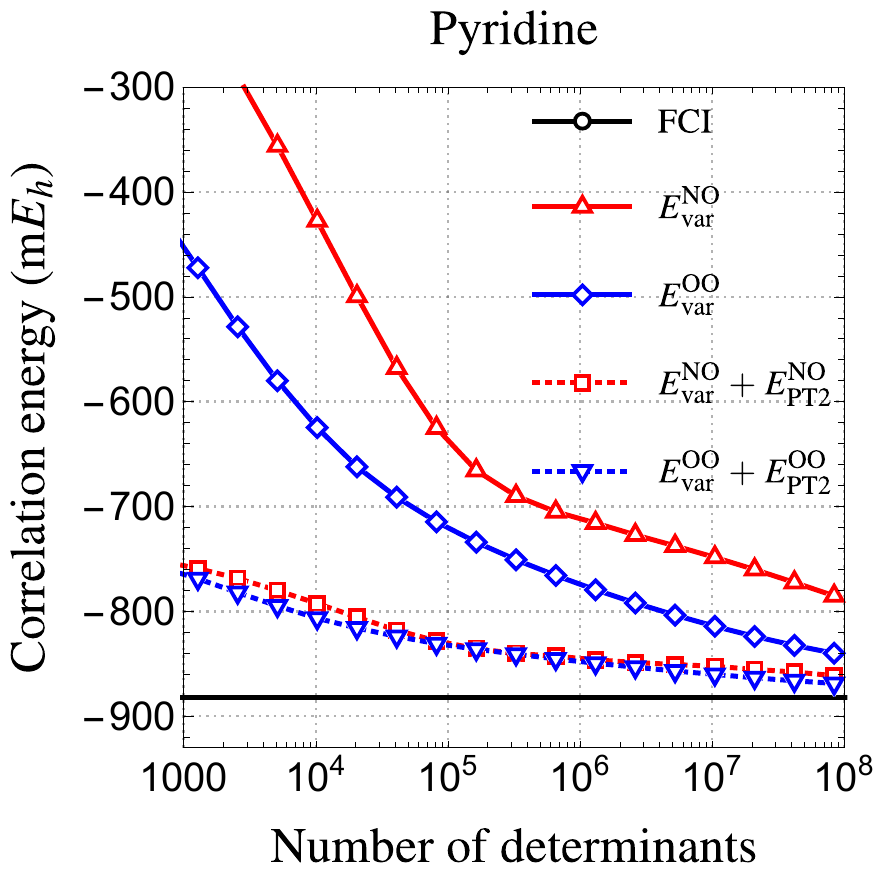}
	\includegraphics[width=0.24\textwidth]{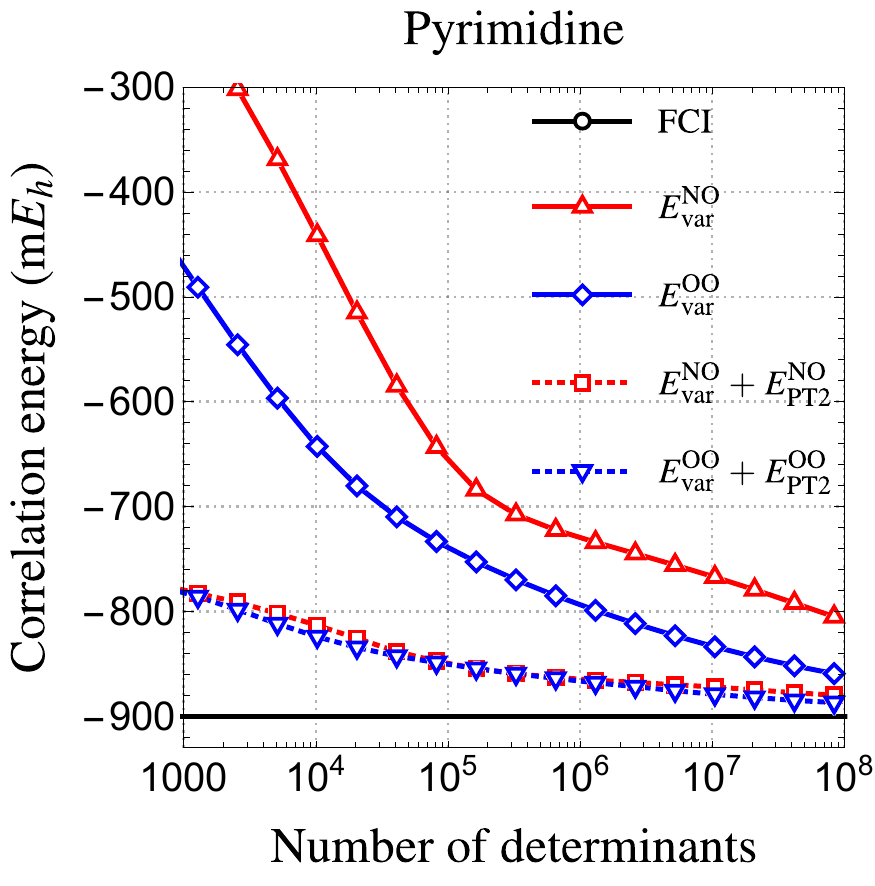}
	\includegraphics[width=0.24\textwidth]{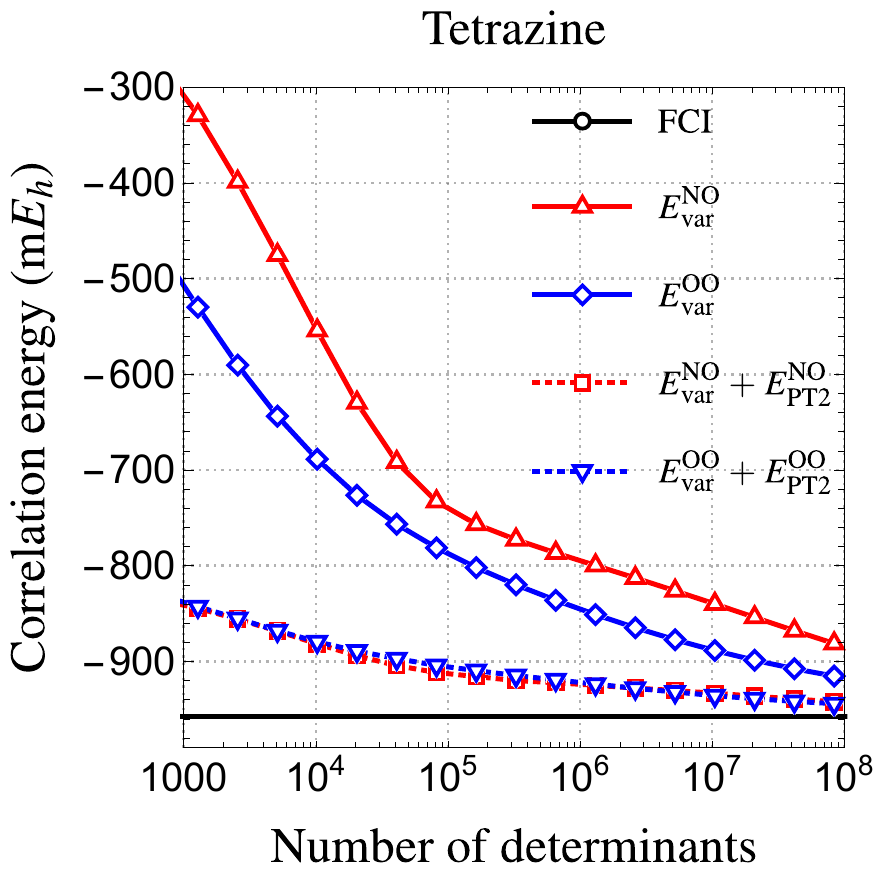}
	\includegraphics[width=0.24\textwidth]{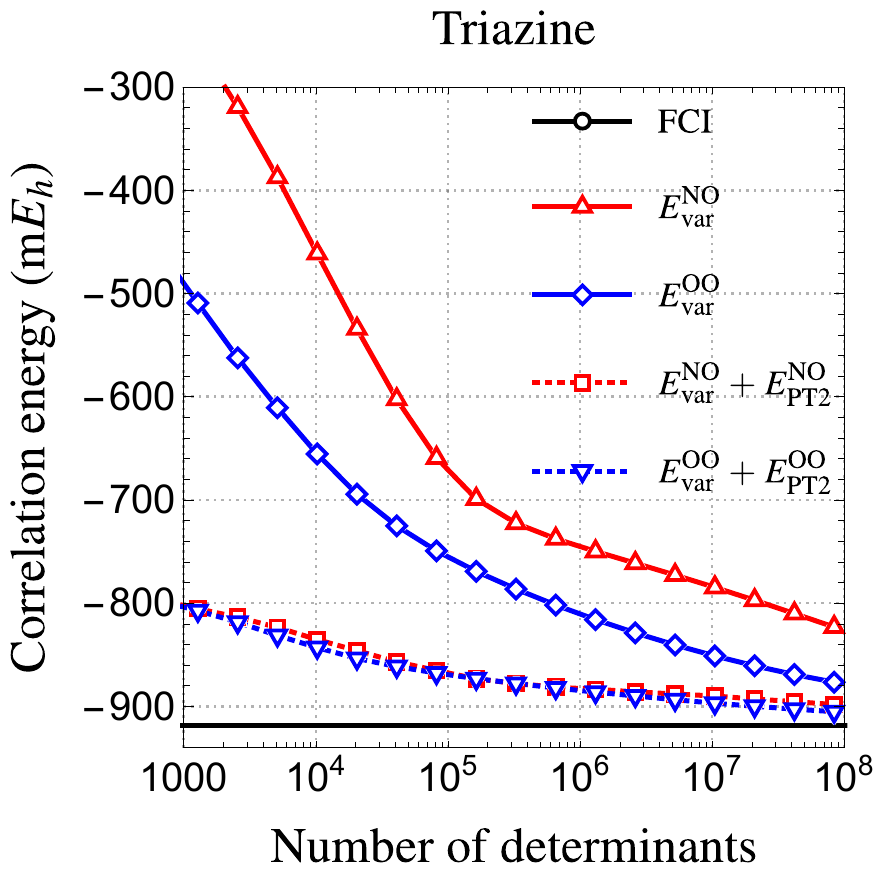}
	\caption{$\Delta \Evar$ (solid) and $\Delta \Evar + \EPT$ (dashed) computed in the cc-pVDZ basis as functions of the number of determinants $\Ndet$ in the variational space for the twelve cyclic molecules represented in Fig.~\ref{fig:mol}.
	Two sets of orbitals are considered: natural orbitals (NOs, in red) and optimized orbitals (OOs, in blue).
	The FCI estimate of the correlation energy is represented as a thick black line.
	\label{fig:vsNdet}
	}
\end{figure*}

\begin{figure*}
	\includegraphics[width=0.24\textwidth]{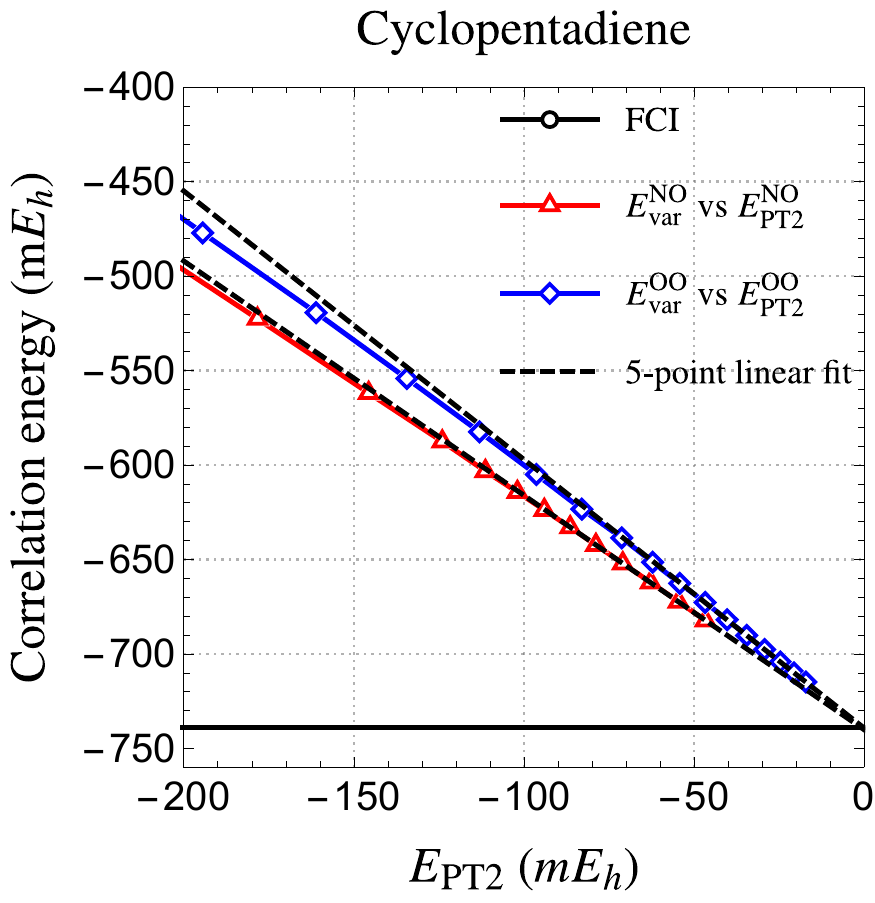}
	\includegraphics[width=0.24\textwidth]{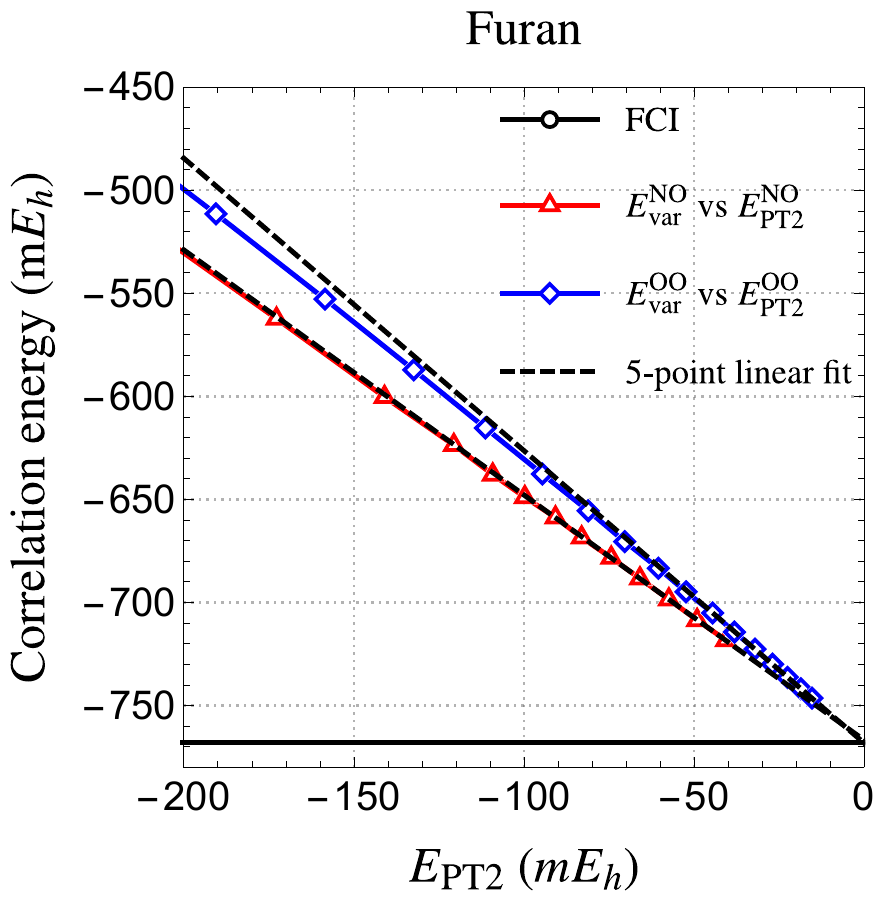}
	\includegraphics[width=0.24\textwidth]{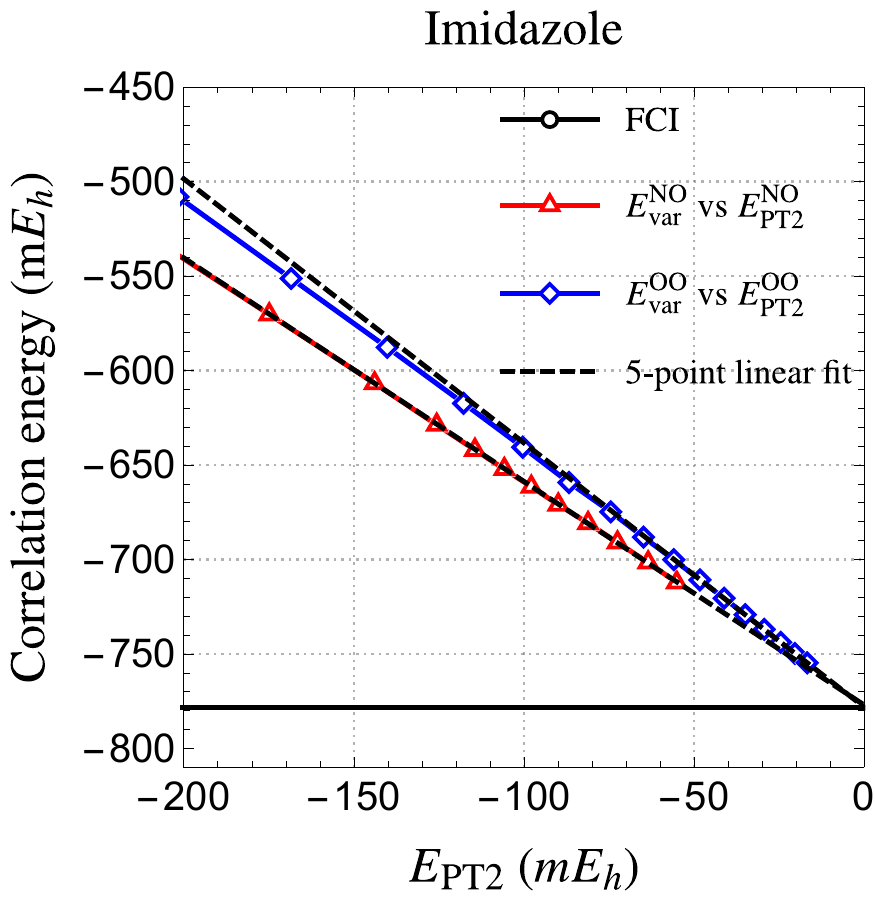}
	\includegraphics[width=0.24\textwidth]{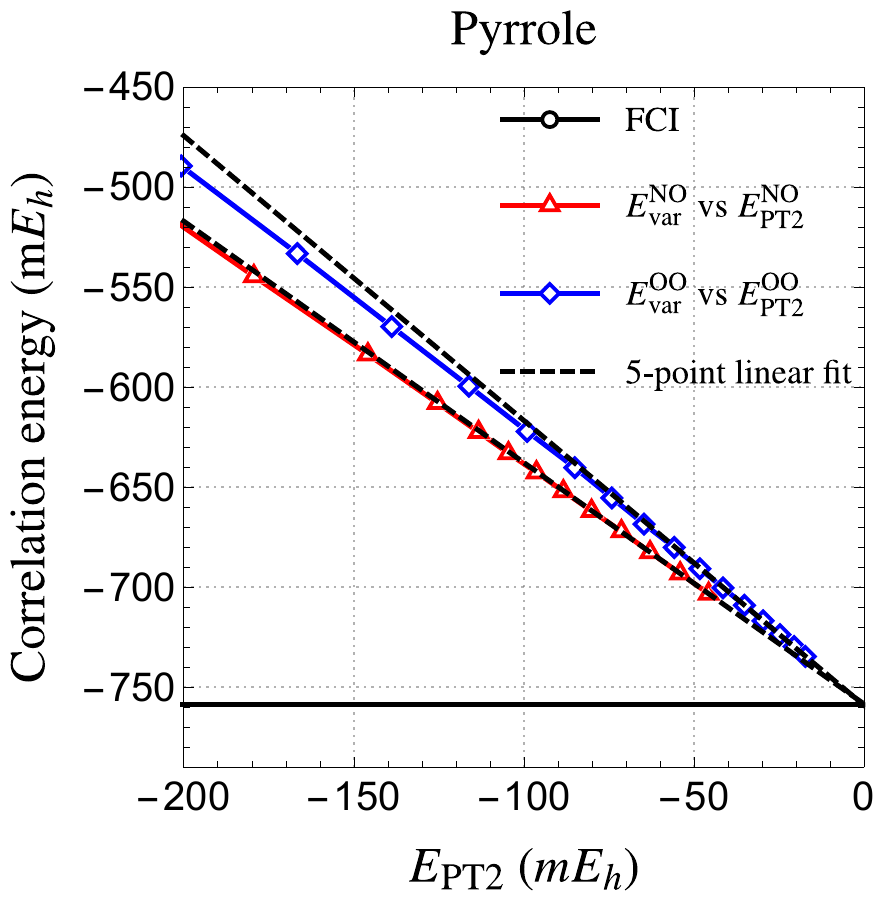}
	\\
	\includegraphics[width=0.24\textwidth]{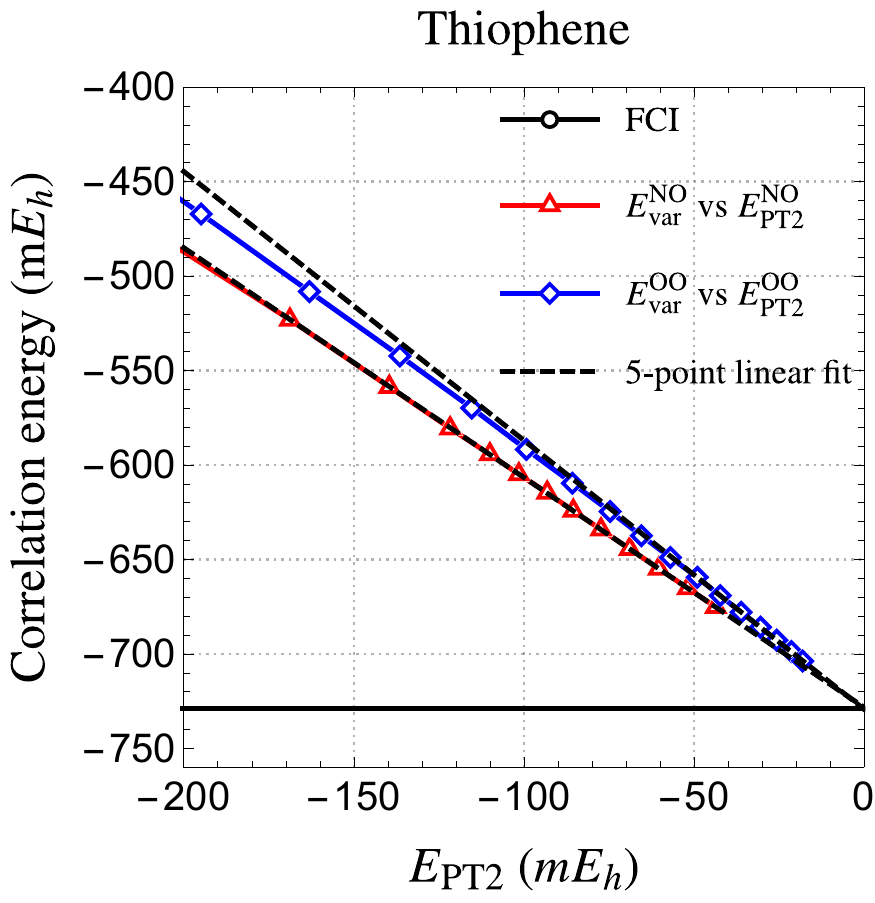}
	\includegraphics[width=0.24\textwidth]{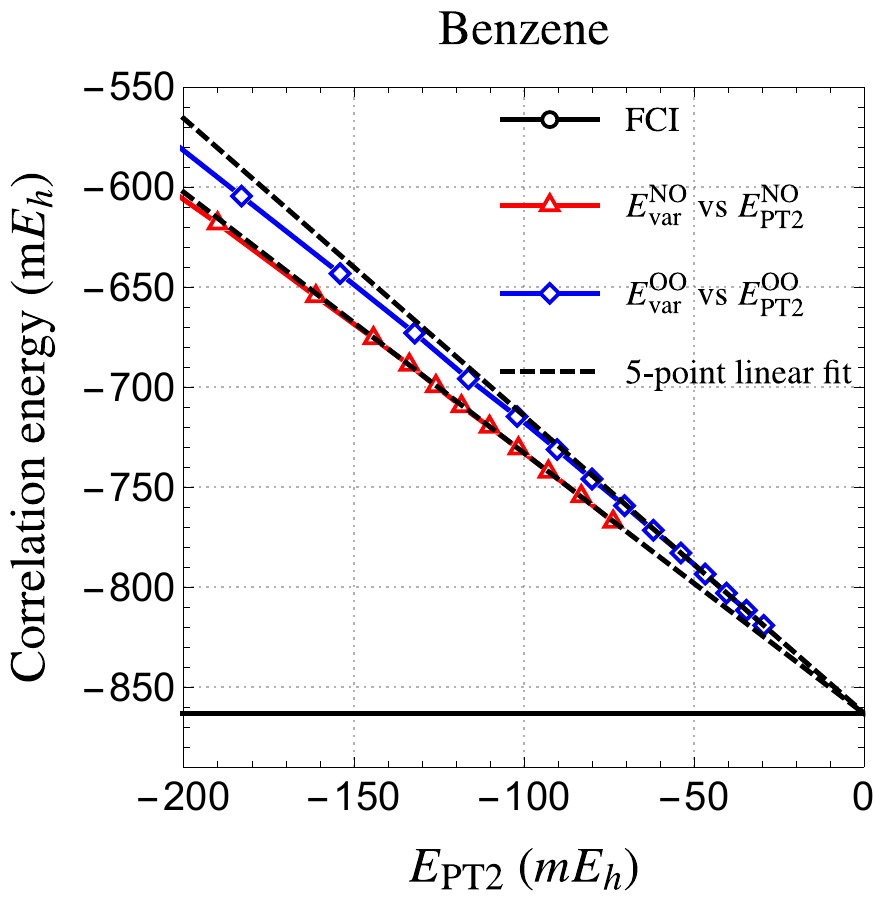}
	\includegraphics[width=0.24\textwidth]{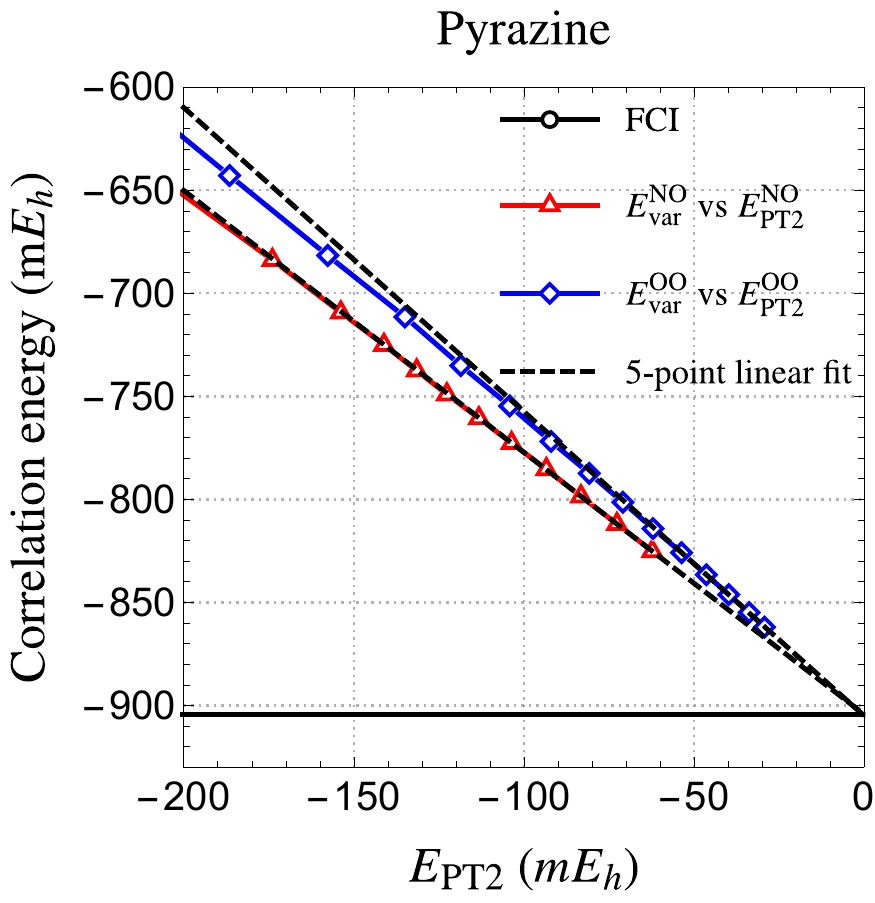}
	\includegraphics[width=0.24\textwidth]{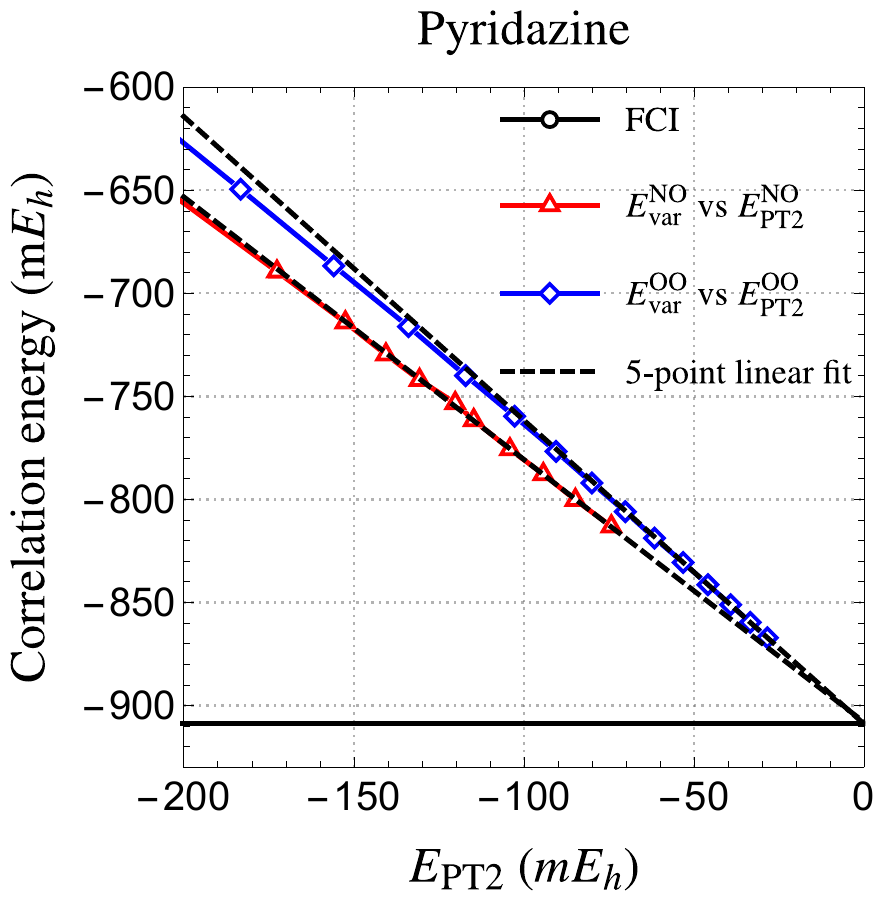}
	\\
	\includegraphics[width=0.24\textwidth]{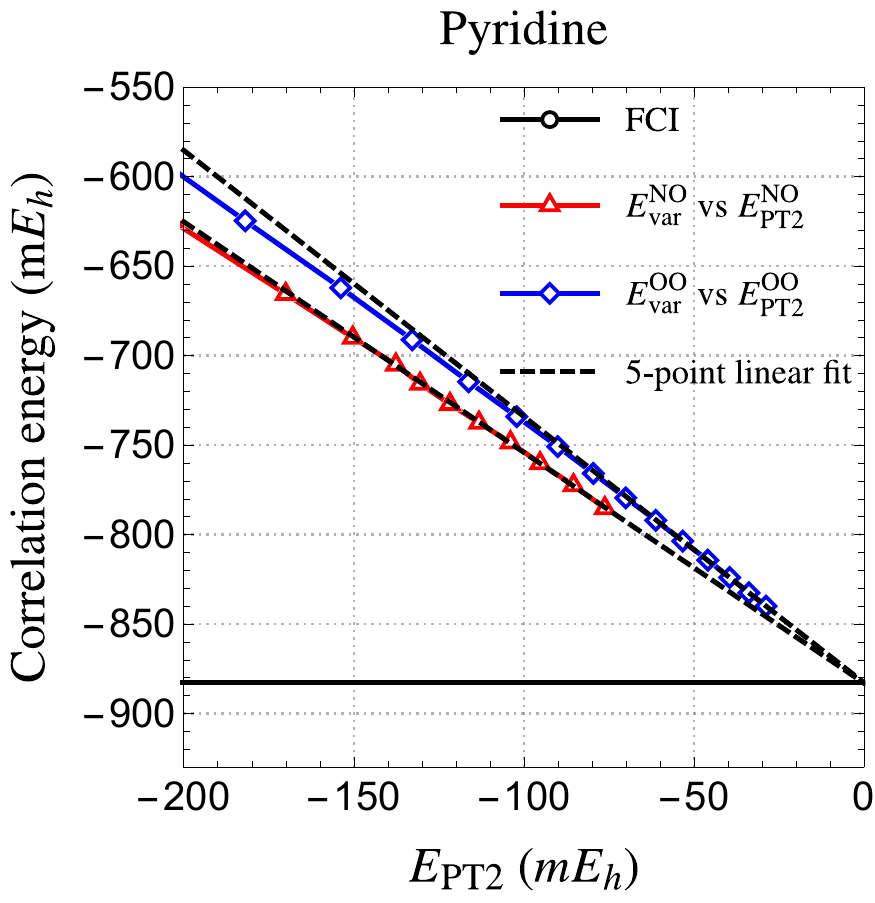}
	\includegraphics[width=0.24\textwidth]{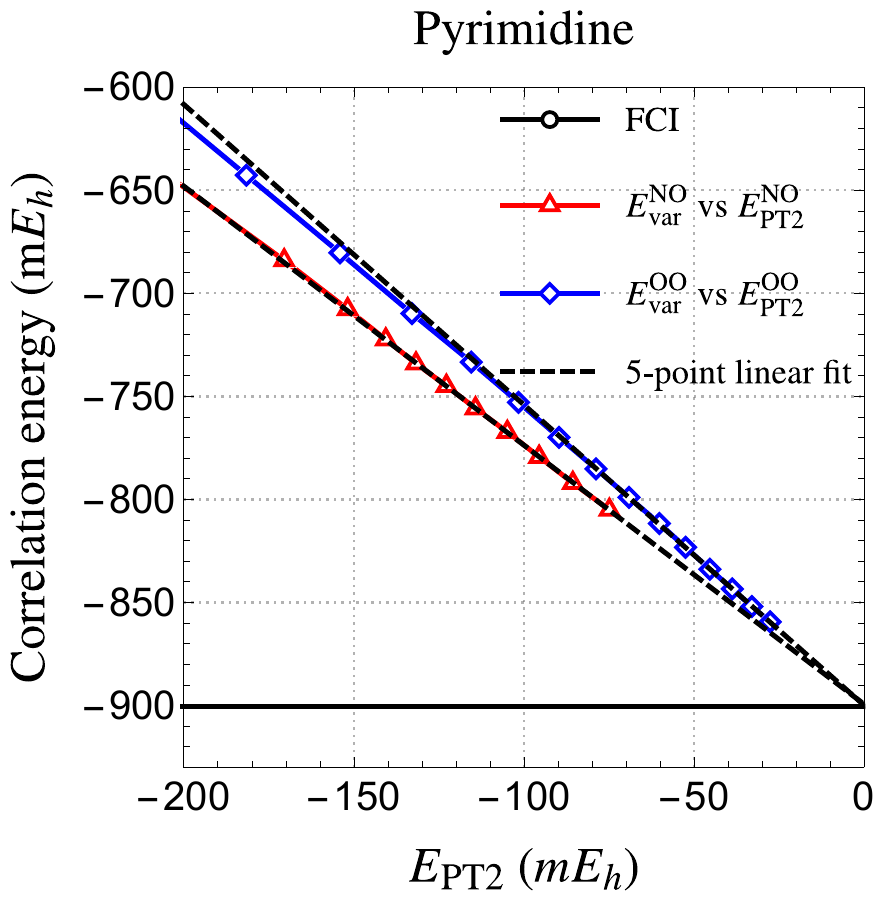}
	\includegraphics[width=0.24\textwidth]{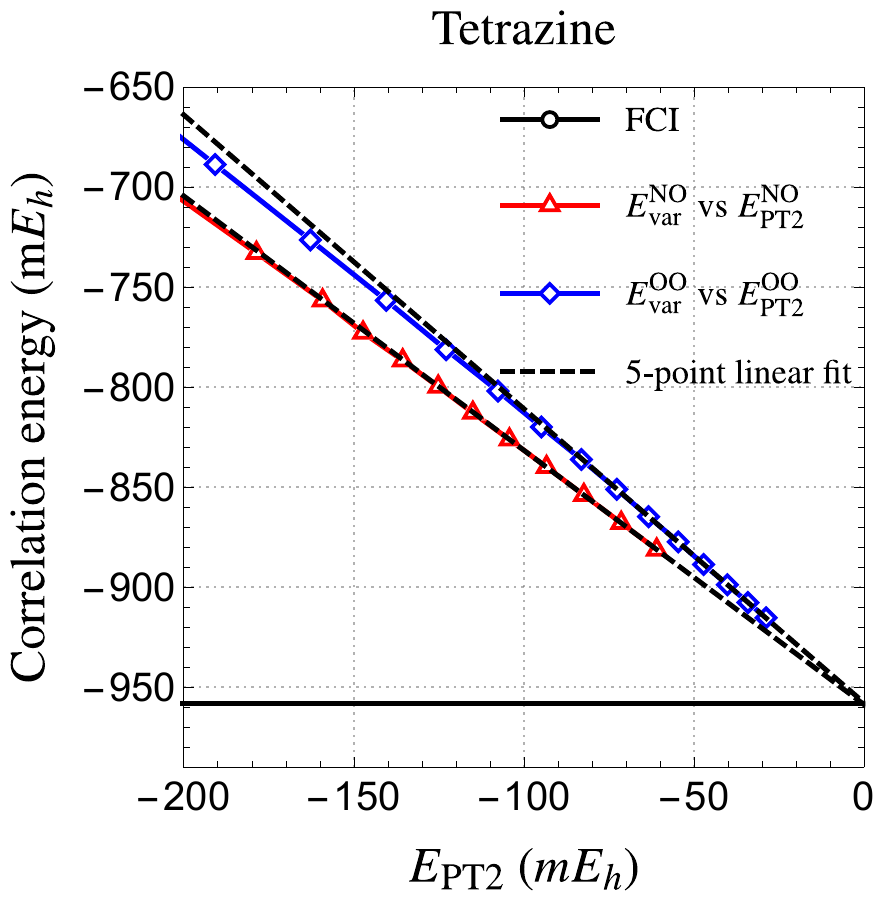}
	\includegraphics[width=0.24\textwidth]{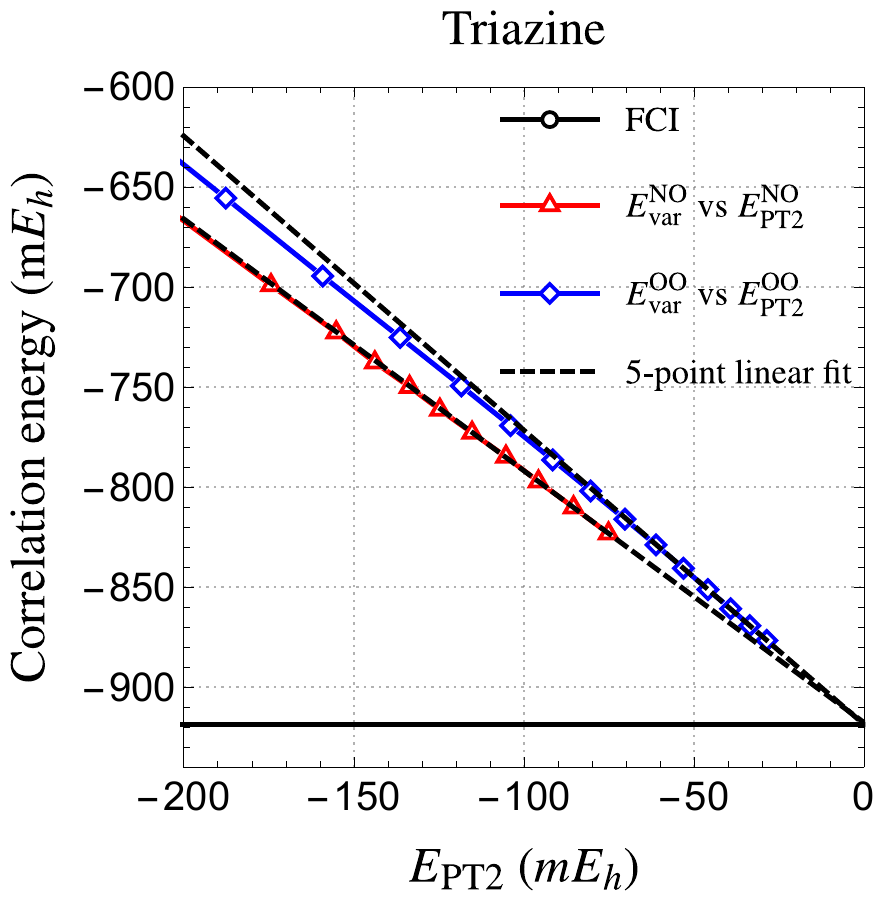}
	\caption{$\Delta \Evar$ as a function of $\EPT$ computed in the cc-pVDZ basis for the twelve cyclic molecules represented in Fig.~\ref{fig:mol}.
	Two sets of orbitals are considered: natural orbitals (NOs, in red) and optimized orbitals (OOs, in blue).
	The five-point weighted linear fit using the five largest variational wave functions for each set is depicted as a dashed black line.
	The weights are taken as the inverse square of the perturbative corrections.
	The FCI estimate of the correlation energy is represented as a thick black line.
	\label{fig:vsEPT2}}
\end{figure*}

\begin{squeezetable}
\begin{table*}
	\caption{Total energy $E$ (in \si{\hartree}) and correlation energy $\Delta E$ (in \si{\milli\hartree}) for the frozen-core ground state of five-membered rings in the cc-pVDZ basis set.
	For the CIPSI estimates of the FCI correlation energy, the fitting error associated with the weighted five-point linear fit is reported in parenthesis.
	\label{tab:Tab5-VDZ}}
	\begin{ruledtabular}
	\begin{tabular}{lcccccccccc}
				&	\mc{2}{c}{Cyclopentadiene}	&	\mc{2}{c}{Furan}	&	\mc{2}{c}{Imidazole}	&	\mc{2}{c}{Pyrrole}	&	\mc{2}{c}{Thiophene}	\\
					\cline{2-3}	\cline{4-5}	\cline{6-7} \cline{8-9} \cline{10-11}
		Method	&	$E$&	$\Delta E$	&	$E$	&	$\Delta E$	&	$E$	&	$\Delta E$	&	$E$	&	$\Delta E$	&	$E$	&	$\Delta E$ 	\\
		\hline
		HF		&	$-192.8083$	&				&	$-228.6433$	&				&	$-224.8354$	&				&	$-208.8286$	&		&$-551.3210$	&	\\
		\hline
		MP2		&	$-193.4717$	&	$-663.4$	&	$-229.3508$	&	$-707.5$	&	$-225.5558$ &	$-720.4$	&	$-209.5243$	&	$-695.7$	&	$-551.9825$	 &	$-661.5$	\\
		MP3		&	$-193.5094$	&	$-701.0$	&	$-229.3711$	&	$-727.8$	&	$-225.5732$	&	$-737.8$	&	$-209.5492$	&	$-720.6$	&	$-552.0104$	&	$-689.4$	\\
		MP4		&	$-193.5428$	&	$-734.5$	&	$-229.4099$	&	$-766.6$	&	$-225.6126$	&	$-777.2$	&	$-209.5851$	&	$-756.5$	&	$-552.0476$	&	$-726.6$	\\
		MP5		&	$-193.5418$	&	$-733.4$	&	$-229.4032$	&	$-759.9$	&	$-225.6061$	&	$-770.8$	&	$-209.5809$	&	$-752.3$	&	$-552.0426$	&	$-721.6$\\
		\hline
		CC2		&	$-193.4782$	&	$-669.9$	&	$-229.3605$	&	$-717.2$	&	$-225.5644$	&	$-729.0$	&	$-209.5311$	&	$-702.5$	&	$-551.9905$	&	$-669.5$	\\
		CC3		&	$-193.5449$	&	$-736.6$	&	$-229.4090$	&	$-765.7$	&	$-225.6115$	&	$-776.1$	&	$-209.5849$	&	$-756.3$	&	$-552.0473$	&	$-726.3$	\\
		CC4		&	$-193.5467$	&	$-738.4$	&	$-229.4102$	&	$-766.9$	&	$-225.6126$	&	$-777.2$	&	$-209.5862$	&	$-757.6$	&	$-552.0487$	&	$-727.7$	\\
		\hline
		CCSD	&	$-193.5156$	&	$-707.2$	&	$-229.3783$	&	$-735.0$	&	$-225.5796$	&	$-744.2$	&	$-209.5543$	&	$-725.7$	&	$-552.0155$	&	$-694.5$	\\
		CCSDT	&	$-193.5446$	&	$-736.2$	&	$-229.4076$	&	$-764.3$	&	$-225.6099$	&	$-774.6$	&	$-209.5838$	&	$-755.2$	&	$-552.0461$	&	$-725.1$	\\
		CCSDTQ	&	$-193.5465$	&	$-738.2$	&	$-229.4100$	&	$-766.7$	&	$-225.6123$	&	$-776.9$	&	$-209.5860$	&	$-757.4$	&	$-552.0485$	&	$-727.5$	\\
		\hline
		CCSD(T)	&	$-193.5439$	&	$-735.6$	&	$-229.4073$	&	$-764.0$	&	$-225.6099$	&	$-774.5$	&	$-209.5836$	&	$-754.9$	&	$-552.0458$	&	$-724.8$
	\\
		CR-CC(2,3)&	$-193.5439$	&	$-735.6$	&	$-229.4075$	&	$-764.2$	&	$-225.6098$	&	$-774.5$	&	$-209.5835$	&	$-754.9$	&	$-552.0459$	&	$-724.9$
	\\
		\hline
		FCI	&					&	$-739.2(1)$	&				&	$-768.2(1)$	&				&	$-778.2(1)$	&				&	$-758.5(1)$	&				&	$-728.9(3)$\\
	\end{tabular}
	\end{ruledtabular}
\end{table*}
\end{squeezetable}

\begin{squeezetable}
\begin{table*}
	\caption{Total energy $E$ (in \si{\hartree}) and correlation energy $\Delta E$ (in \si{\milli\hartree}) for the frozen-core ground state of six-membered rings in the cc-pVDZ basis set.
	For the CIPSI estimates of the FCI correlation energy, the fitting error associated with the weighted five-point linear fit is reported in parenthesis.
	\label{tab:Tab6-VDZ}}
	\begin{ruledtabular}
	\begin{tabular}{lcccccccccccccc}
				&	\mc{2}{c}{Benzene}	&	\mc{2}{c}{Pyrazine}	&	\mc{2}{c}{Pyridazine}	&	\mc{2}{c}{Pyridine}	&	\mc{2}{c}{Pyrimidine}	&	\mc{2}{c}{s-Tetrazine}	&	\mc{2}{c}{s-Triazine}	\\
					\cline{2-3}	\cline{4-5}	\cline{6-7} \cline{8-9} \cline{10-11} \cline{12-13} \cline{14-15}
	Method		&	$E$	&	$\Delta E$	&	$E$	&	$\Delta E$	&	$E$	&	$\Delta E$	&	$E$	&	$\Delta E$
				&	$E$	&	$\Delta E$	&	$E$	&	$\Delta E$	&	$E$	&	$\Delta E$	\\
	\hline
		HF		&	$-230.7222$	&			&	$-262.7030$	&			&	$-262.6699$	&			&	$-246.7152$	&			&	$-262.7137$	&			&	$-294.6157$		&		&	$-278.7173$	\\
		\hline
		MP2		&	$-231.5046$  	&	$-782.3$	&	$-263.5376$ 	&	$-834.6$ 	&	$-263.5086$ &	$-838.7$	&	$-247.5227$	&	$-807.5$	&	$-263.5437$   &	$-830.1$	&	$-295.5117$    &   $-895.9$	&			$-279.5678$   &   $-850.5$\\
		MP3		&	$-231.5386$	&	$-816.4$	&	$-263.5567$	&	$-853.7$	&	$-263.5271$	&	$-857.3$	&	$-247.5492$	&	$-834.0$	&	$-263.5633$	&	$-849.6$	&	$-295.5152$   &	$-899.5$	&	$-279.5809$   &	$-863.6$	\\
		MP4		&	$-231.5808$	&	$-858.5$	&	$-263.6059$	&	$-902.9$	&	$-263.5778$	&	$-907.9$	&	$-247.5951$	&	$-879.9$	&	$-263.6129$	&	$-899.3$	&	$-295.5743$	&  $-958.6$	&	$-279.6340$    &	$-916.7$	\\
		MP5		&	$-231.5760$   &          $-853.8$	&	$-263.5968$	&	$-893.8$	&	$-263.5681$	&	$-898.3$	&	$-247.5881$	&	$-872.9$	&	$-263.6036$ &            $-890.0$	&	$-295.5600$	&	$-944.3$	&	$-279.6228$	&	$-905.4$	\\
		\hline
		CC2		&	$-231.5117$  	&	$-789.4$	&	$-263.5475$	&	$-844.5$	&	$-263.5188$	&	$-848.9$	&	$-247.5315$   &	$-816.3$	&	$-263.5550$ 	&	$-841.3$	&	$-295.5247$    &   $-909.0$		&	$-279.5817$       &	$-864.4$	\\
		CC3		&	$-231.5814$	&	$-859.1$	&	$-263.6045$	&	$-901.5$	&	$-263.5761$	&	$-906.2$	&	$-247.5948$	&	$-879.6$	&	$-263.6120$	&	$-898.4$	&	$-295.5706$  &	$-954.9$	&	$-279.6329$    &	$-915.6$	\\
		CC4		&	$-231.5828$	&	$-860.6$	&	$-263.6056$	&	$-902.6$	&	$-263.5773$	&	$-907.5$	&	$-247.5960$		&	$-880.8$		&	$-263.6129$	&	$-899.3$		&	$-295.5716$	&	$-955.9$	&	$-279.6334$	&	$-916.1$			\\
		\hline
		CCSD	&	$-231.5440$	&	$-821.8$	&	$-263.5640$	&	$-861.0$	&	$-263.5347$	&	$-864.9$	&	$-247.5559$	&	$-840.7$	&	$-263.5716$	&	$-858.0$	&	$-295.5248$   &	$-909.1$	&	$-279.5911$   &	 $-873.8$	\\
		CCSDT	&	$-231.5802$	&	$-857.9$	&	$-263.6024$	&	$-899.4$	&	$-263.5739$	&	$-904.0$	&	$-247.5931$	&	$-877.9$	&	$-263.6097$	&	$-896.1$	&	$-295.5673$  &  $-951.6$	&	$-279.6300$    &	$-912.7$	\\
		CCSDTQ	&	$-231.5826$	&	$-860.4$		&	$-263.6053$	&	$-902.3$		&	$-263.5770$	&	$-907.1$	&	$-247.5960$	&	$-880.8$	&	$-263.6126$  & $-899.0$		&			$-295.5712$		&	$-955.4$		&	$-279.6331$		&	$-915.8$		\\
		\hline
		CCSD(T)	&	$-231.5798$	&	$-857.5$	&	$-263.6024$		&	$-899.4$	&	$-263.5740$	&	$-904.1$	&	$-247.5929$	&	$-877.7$	&	$-263.6099$	&	$-896.2$	&	$-295.5680$		&	$-952.2$	&	$-279.6305$		&	$-913.1$	\\
		CR-CC(2,3)	&	$-231.5792$	&	$-856.9$	&	$-263.6017$		&	$-898.7$	&	$-263.5732$	&	$-903.3$	&	$-247.5922$	&	$-877.1$	&	$-263.6091$	&	$-895.5$	&	$-295.5670$		&	$-951.3$	&	$-279.6298$		&	$-912.5$	\\
		\hline
		FCI	&				&	$-862.9(3)$		&				&	$-904.6(4)$		&				&	$-908.8(1)$		&				&	$-882.7(4)$		&				&	$-900.5(2)$		&		&	$-957.9(4)$		&		&	$-918.4(3)$\\
	\end{tabular}
	\end{ruledtabular}
\end{table*}
\end{squeezetable}

\begin{squeezetable}
\begin{table*}
	\caption{
	\alert{Extrapolation distance $\Delta \Edist$ (in \si{\milli\hartree}) defined as the difference between the final computed energy $\Delta \Efinal$ (in \si{\milli\hartree}) and the extrapolated correlation energies $\Delta \Eextrap$ (in \si{\milli\hartree}) computed in the cc-pVDZ basis for the twelve cyclic molecules represented in Fig.~\ref{fig:mol} and their associated fitting errors (in \si{\milli\hartree}) obtained via weighted linear fits with a varying number of points.}
	Two sets of orbitals are considered: natural orbitals and optimized orbitals.
	The weights are taken as the inverse square of the perturbative corrections.
	For a $m$-point fit, the $m$ largest variational wave functions are used.
	\label{tab:fit}}
	\begin{ruledtabular}
	\begin{tabular}{lccccccccc}
	Molecule	&	Number of		&	\mc{4}{c}{Natural orbitals}	&	\mc{4}{c}{Optimized orbitals}	\\
									\cline{3-6}\cline{7-10}
				&	 fitting points	&	$\Delta \Efinal$	&	$\Delta \Eextrap$	&	$\Delta \Edist$	&	Fitting error	
									&	$\Delta \Efinal$	&	$\Delta \Eextrap$	&	$\Delta \Edist$	&	Fitting error		\\
	\hline
	Cyclopentadiene	&	3	&	$-728.941$	&	$-740.639$	&	$11.699$	&	$0.273$	&	$-731.987$ 	&	$-739.295$	&	$7.308$	&	$0.199$		\\
					&	4	&	$-728.941$	&	$-740.243$	&	$11.303$	&	$0.306$	&	$-731.987$ 	&	$-739.309$	&	$7.322$	&	$0.088$		\\
					&\bf5	&	$-728.941$	&	$-740.047$	&	$11.106$	&	$0.242$	&$\bf-731.987$	&$\bf-739.230$	&$\bf7.243$	&$\bf0.074$		\\
					&	6	&	$-728.941$	&	$-739.952$	&	$11.011$	&	$0.187$	&	$-731.987$ 	&	$-739.304$	&	$7.317$	&	$0.072$		\\
					&	7	&	$-728.941$	&	$-739.761$	&	$10.820$	&	$0.204$	&	$-731.987$ 	&	$-739.292$	&	$7.305$	&	$0.055$		\\
	\hline
	Furan			&	3	&	$-758.946$	&	$-766.090$	&	$7.144$		&	$0.729$	&	$-761.715$	&	$-767.790$	&	$6.076$	&	$0.064$		\\
					&	4	&	$-758.946$	&	$-766.445$	&	$7.499$		&	$0.459$	&	$-761.715$	&	$-768.104$	&	$6.389$	&	$0.196$		\\
					&\bf5	&	$-758.946$	&	$-766.582$	&	$7.636$		&	$0.318$	&$\bf-761.715$	&$\bf-768.194$	&$\bf6.479$	&$\bf0.135$		\\
					&	6	&	$-758.946$	&	$-766.366$	&	$7.420$		&	$0.288$	&	$-761.715$	&	$-768.060$	&	$6.345$	&	$0.131$		\\
					&	7	&	$-758.946$	&	$-766.507$	&	$7.561$		&	$0.254$	&	$-761.715$	&	$-768.086$	&	$6.372$	&	$0.101$		\\
	\hline
	Imidazole		&	3	&	$-767.314$	&	$-778.148$	&	$10.833$	&	$2.197$	&	$-771.362$	&	$-778.295$	&	$6.932$	&	$0.356$		\\
					&	4	&	$-767.314$	&	$-777.436$	&	$10.122$	&	$1.107$	&	$-771.362$	&	$-778.270$	&	$6.908$	&	$0.150$		\\
					&\bf5	&	$-767.314$	&	$-776.300$	&	$8.986$		&	$0.996$	&$\bf-771.362$	&$\bf-778.178$	&$\bf6.816$	&$\bf0.105$		\\
					&	6	&	$-767.314$	&	$-776.104$	&	$8.789$		&	$0.712$	&	$-771.362$	&	$-778.174$	&	$6.812$	&	$0.072$		\\
					&	7	&	$-767.314$	&	$-776.098$	&	$8.784$		&	$0.541$	&	$-771.362$	&	$-778.051$	&	$6.689$	&	$0.099$		\\
	\hline
	Pyrrole			&	3	&	$-748.961$	&	$-758.309$	&	$9.348$		&	$0.447$	&	$-751.862$	&	$-758.650$	&	$6.788$	&	$0.321$		\\
					&	4	&	$-748.961$	&	$-758.749$	&	$9.788$		&	$0.393$	&	$-751.862$	&	$-758.389$	&	$6.527$	&	$0.174$		\\
					&\bf5	&	$-748.961$	&	$-758.405$	&	$9.444$		&	$0.359$	&$\bf-751.862$	&$\bf-758.460$	&$\bf6.598$	&$\bf0.110$		\\
					&	6	&	$-748.961$	&	$-758.136$	&	$9.175$		&	$0.334$	&	$-751.862$	&	$-758.352$	&	$6.490$	&	$0.100$		\\
					&	7	&	$-748.961$	&	$-757.990$	&	$9.029$		&	$0.283$	&	$-751.862$	&	$-758.347$	&	$6.485$	&	$0.075$		\\
	\hline
	Thiophene		&	3	&	$-718.769$	&	$-728.054$	&	$9.285$		&	$0.134$	&	$-721.757$	&	$-728.744$	&	$6.987$	&	$0.691$		\\
					&	4	&	$-718.769$	&	$-728.240$	&	$9.471$		&	$0.139$	&	$-721.757$	&	$-729.052$	&	$7.295$	&	$0.331$		\\
					&\bf5	&	$-718.769$	&	$-728.243$	&	$9.474$		&	$0.087$	&$\bf-721.757$	&$\bf-728.948$	&$\bf7.191$	&$\bf0.203$		\\
					&	6	&	$-718.769$	&	$-728.242$	&	$9.472$		&	$0.062$	&	$-721.757$	&	$-728.987$	&	$7.230$	&	$0.140$		\\
					&	7	&	$-718.769$	&	$-728.420$	&	$9.651$		&	$0.144$	&	$-721.757$	&	$-729.067$	&	$7.310$	&	$0.117$		\\
	\hline
	Benzene			&	3	&	$-841.030$	&	$-860.350$	&	$19.3197$	&	$0.496$	&	$-848.540$	&	$-862.325$	&	$13.7847$	&	$0.279$		\\
					&	4	&	$-841.030$	&	$-861.949$	&	$20.9186$	&	$0.811$	&	$-848.540$	&	$-863.024$	&	$14.4842$	&	$0.424$		\\
					&\bf5	&	$-841.030$	&	$-861.807$	&	$20.7772$	&	$0.474$	&$\bf-848.540$	&$\bf-862.890$	&$\bf14.3496$	&$\bf0.266$		\\
					&	6	&	$-841.030$	&	$-861.110$	&	$20.0803$	&	$0.539$	&	$-848.540$	&	$-862.360$	&	$13.8202$	&	$0.383$		\\
					&	7	&	$-841.030$	&	$-861.410$	&	$20.3794$	&	$0.444$	&	$-848.540$	&	$-862.083$	&	$13.5435$	&	$0.339$		\\
	\hline
	Pyrazine		&	3	&	$-887.414$	&	$-904.148$	&	$16.734$	&	$0.035$	&	$-891.249$	&	$-904.867$	&	$13.619$	&	$1.420$		\\
					&	4	&	$-887.414$	&	$-904.726$	&	$17.312$	&	$0.377$	&	$-891.249$	&	$-904.588$	&	$13.340$	&	$0.650$		\\
					&\bf5	&	$-887.414$	&	$-904.274$	&	$16.859$	&	$0.383$	&$\bf-891.249$	&$\bf-904.550$	&$\bf13.301$	&$\bf0.385$		\\
					&	6	&	$-887.414$	&	$-903.980$	&	$16.566$	&	$0.341$	&	$-891.249$	&	$-903.982$	&	$12.734$	&	$0.439$		\\
					&	7	&	$-887.414$	&	$-903.621$	&	$16.206$	&	$0.370$	&	$-891.249$	&	$-903.746$	&	$12.497$	&	$0.359$		\\
	\hline
	Pyridazine		&	3	&	$-887.410$	&	$-910.856$	&	$23.446$	&	$3.053$	&	$-895.565$	&	$-909.292$	&	$13.726$	&	$0.024$		\\
					&	4	&	$-887.410$	&	$-908.222$	&	$20.811$	&	$1.834$	&	$-895.565$	&	$-908.808$	&	$13.243$	&	$0.230$		\\
					&\bf5	&	$-887.410$	&	$-909.282$	&	$21.871$	&	$1.191$	&$\bf-895.565$	&$\bf-908.820$	&$\bf13.255$	&$\bf0.133$		\\
					&	6	&	$-887.410$	&	$-912.566$	&	$25.156$	&	$1.727$	&	$-895.565$	&	$-908.342$	&	$12.777$	&	$0.303$		\\
					&	7	&	$-887.410$	&	$-910.694$	&	$23.283$	&	$2.210$	&	$-895.565$	&	$-908.368$	&	$12.802$	&	$0.224$		\\
	\hline
	Pyridine		&	3	&	$-861.424$	&	$-883.025$	&	$21.601$	&	$3.919$	&	$-868.803$	&	$-883.363$	&	$14.560$	&	$0.047$		\\
					&	4	&	$-861.424$	&	$-883.862$	&	$22.438$	&	$1.869$	&	$-868.803$	&	$-883.413$	&	$14.610$	&	$0.029$		\\
					&\bf5	&	$-861.424$	&	$-881.664$	&	$20.240$	&	$1.760$	&$\bf-868.803$	&$\bf-882.700$	&$\bf13.897$	&$\bf0.405$		\\
					&	6	&	$-861.424$	&	$-880.422$	&	$18.998$	&	$1.456$	&	$-868.803$	&	$-882.361$	&	$13.558$	&	$0.341$		\\
					&	7	&	$-861.424$	&	$-880.191$	&	$18.768$	&	$1.084$	&	$-868.803$	&	$-882.023$	&	$13.221$	&	$0.330$		\\
	\hline
	Pyrimidine		&	3	&	$-879.958$	&	$-900.386$	&	$20.428$	&	$1.884$	&	$-887.009$	&	$-900.817$	&	$13.808$	&	$0.726$		\\
					&	4	&	$-879.958$	&	$-901.441$	&	$21.483$	&	$0.991$	&	$-887.009$	&	$-900.383$	&	$13.374$	&	$0.356$		\\
					&\bf5	&	$-879.958$	&	$-900.354$	&	$20.396$	&	$0.865$	&$\bf-887.009$	&$\bf-900.496$	&$\bf13.487$	&$\bf0.214$		\\
					&	6	&	$-879.958$	&	$-900.240$	&	$20.283$	&	$0.594$	&	$-887.009$	&	$-900.698$	&	$13.689$	&	$0.190$		\\
					&	7	&	$-879.958$	&	$-899.689$	&	$19.732$	&	$0.565$	&	$-887.009$	&	$-900.464$	&	$13.455$	&	$0.206$		\\
	\hline
	s-Tetrazine		&	3	&	$-942.162$	&	$-958.736$	&	$16.574$	&	$0.320$	&	$-944.077$	&	$-957.559$	&	$13.4815$	&	$0.246$		\\
					&	4	&	$-942.162$	&	$-958.727$	&	$16.564$	&	$0.148$	&	$-944.077$	&	$-957.299$	&	$13.2221$	&	$0.160$		\\
					&\bf5	&	$-942.162$	&	$-958.500$	&	$16.337$	&	$0.172$	&$\bf-944.077$	&$\bf-957.869$	&$\bf13.7916$	&$\bf0.349$		\\
					&	6	&	$-942.162$	&	$-958.162$	&	$16.000$	&	$0.260$	&	$-944.077$	&	$-957.744$	&	$13.6665$	&	$0.247$		\\
					&	7	&	$-942.162$	&	$-958.161$	&	$15.999$	&	$0.198$	&	$-944.077$	&	$-957.709$	&	$13.6319$	&	$0.183$		\\
	\hline
	s-Triazine		&	3	&	$-898.283$	&	$-917.221$	&	$18.938$	&	$0.693$	&	$-905.180$	&	$-919.596$	&	$14.4152$	&	$0.105$		\\
					&	4	&	$-898.283$	&	$-918.723$	&	$20.440$	&	$0.913$	&	$-905.180$	&	$-918.457$	&	$13.2768$	&	$0.538$		\\
					&\bf5	&	$-898.283$	&	$-917.402$	&	$19.119$	&	$0.956$	&$\bf-905.180$	&$\bf-918.355$	&$\bf13.1745$	&$\bf0.312$		\\
					&	6	&	$-898.283$	&	$-916.517$	&	$18.233$	&	$0.862$	&	$-905.180$	&	$-918.206$	&	$13.0251$	&	$0.226$		\\
					&	7	&	$-898.283$	&	$-916.544$	&	$18.261$	&	$0.643$	&	$-905.180$	&	$-917.876$	&	$12.6956$	&	$0.267$		\\
	\end{tabular}
	\end{ruledtabular}
\end{table*}
\end{squeezetable}

\begin{figure}
	\includegraphics[width=\linewidth]{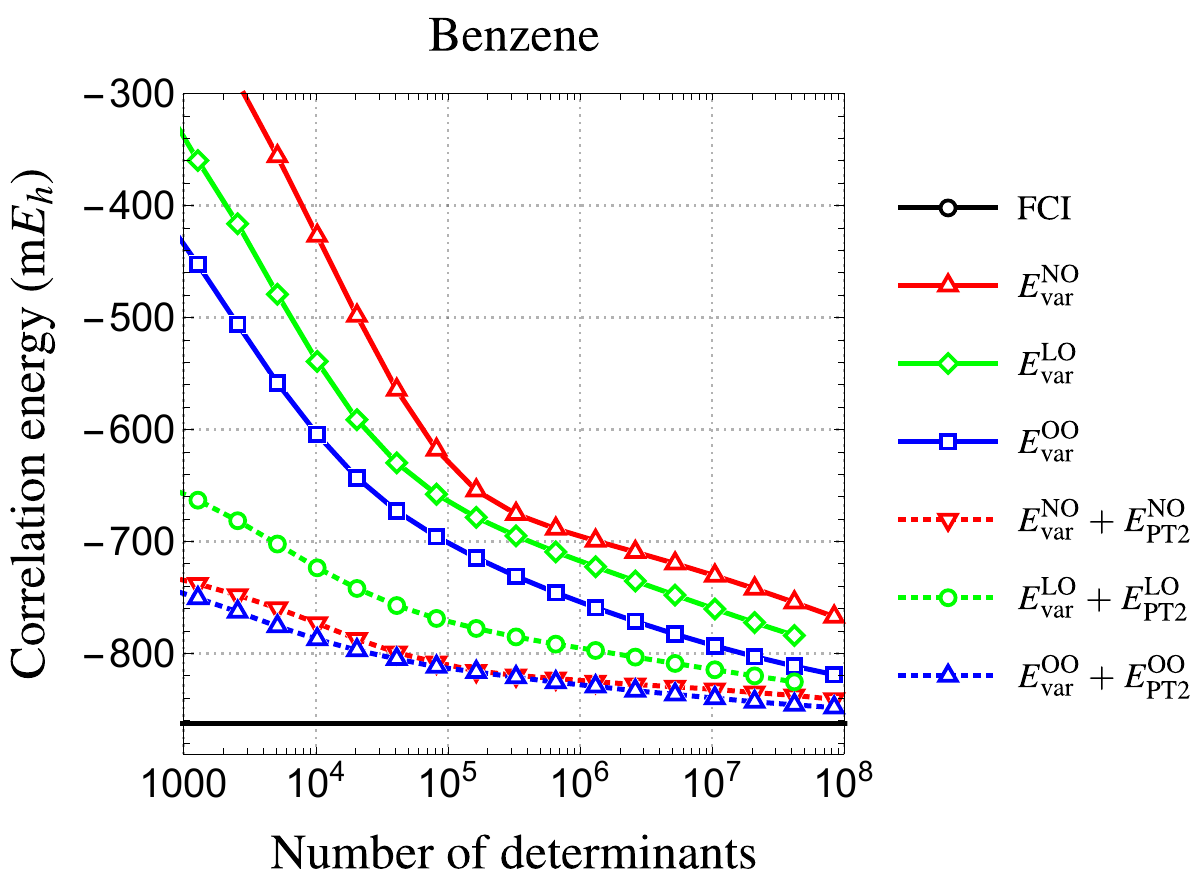}
	\caption{$\Delta \Evar$ (solid) and $\Delta \Evar + \EPT$ (dashed) computed in the cc-pVDZ basis as functions of the number of determinants $\Ndet$ in the variational space for the benzene molecule.
	Three sets of orbitals are considered: natural orbitals (NOs, in red), localized orbitals (LOs, in green), and optimized orbitals (OOs, in blue).
	The FCI estimate of the correlation energy is represented as a thick black line.
	\label{fig:BenzenevsNdet}}
\end{figure}

\begin{figure*}
	\includegraphics[width=0.24\textwidth]{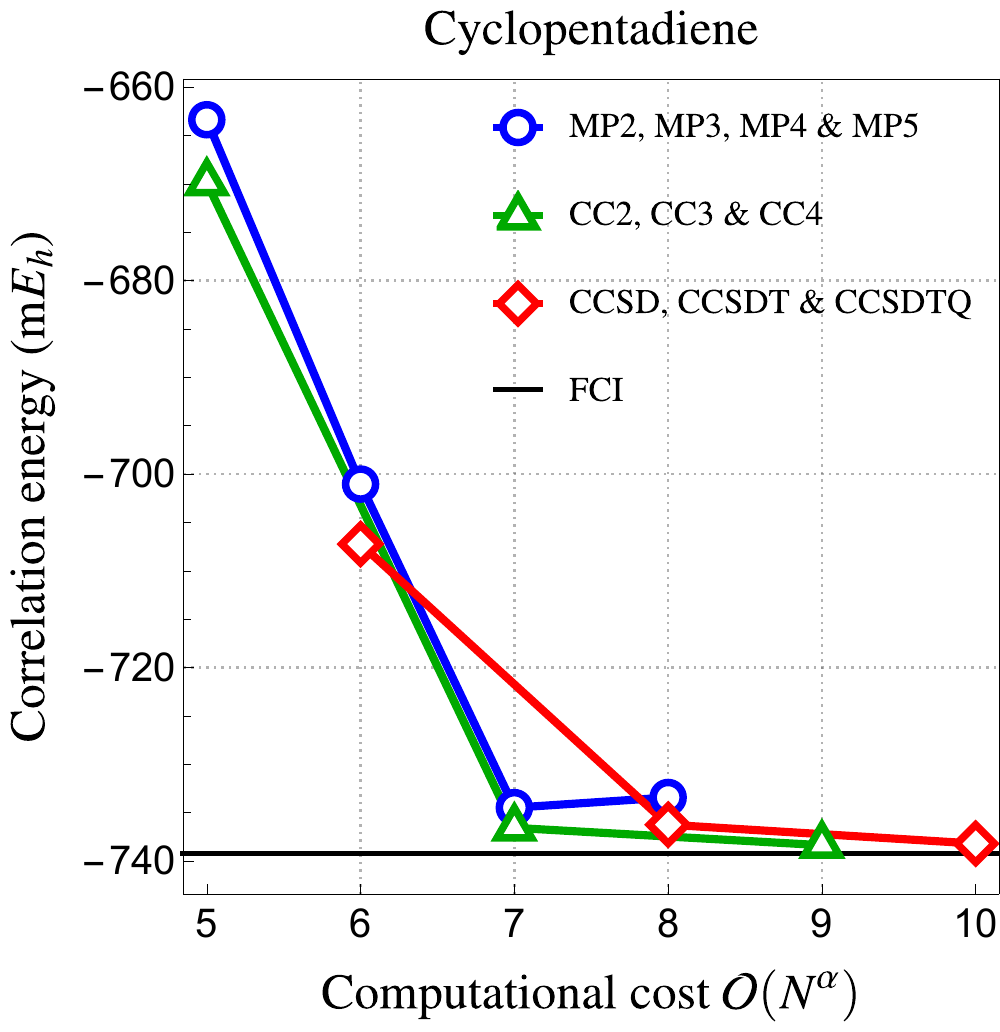}
	\includegraphics[width=0.24\textwidth]{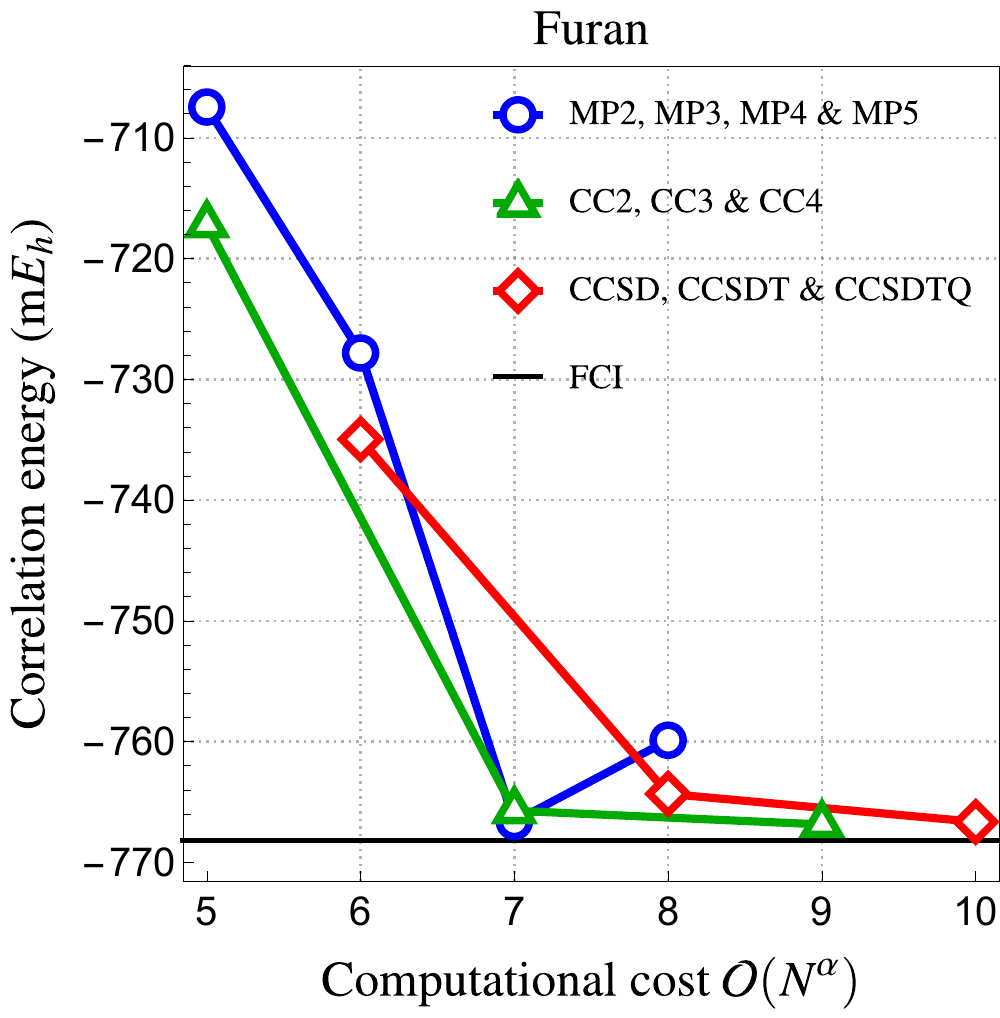}
	\includegraphics[width=0.24\textwidth]{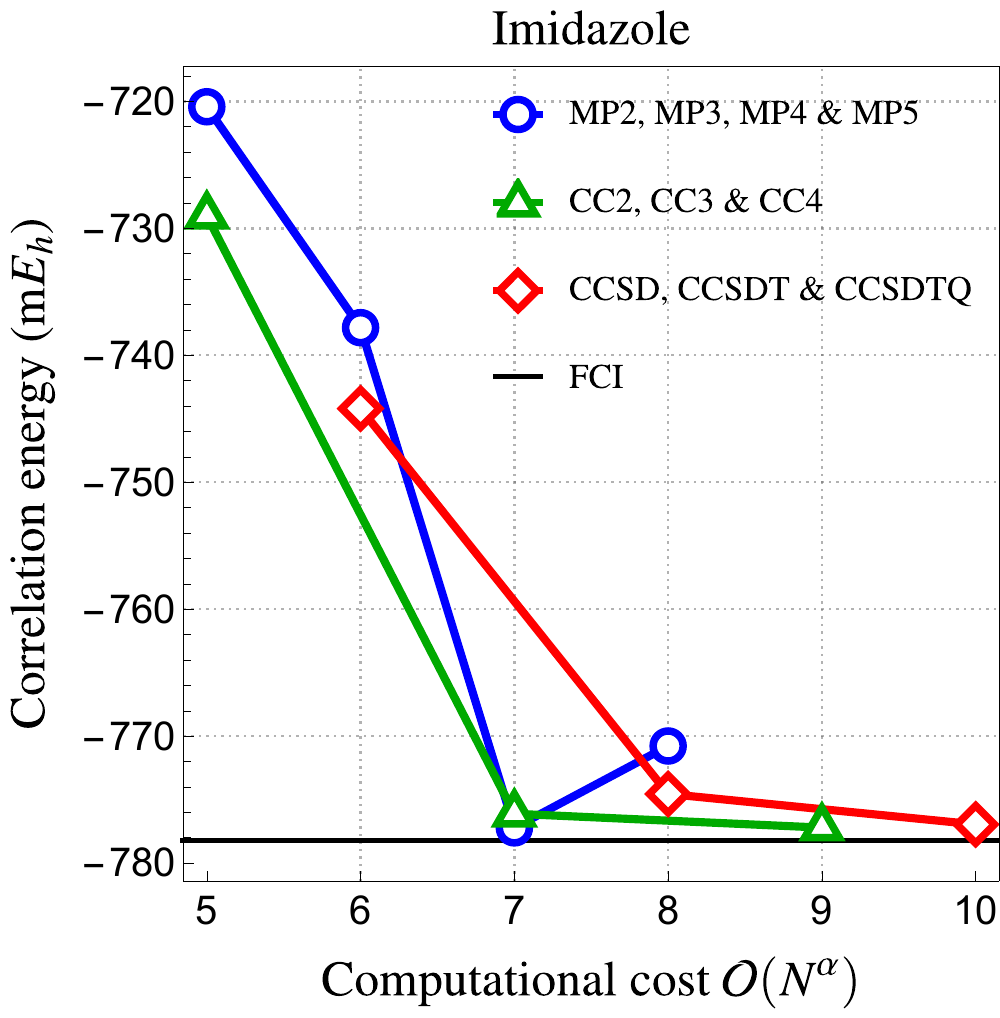}
	\includegraphics[width=0.24\textwidth]{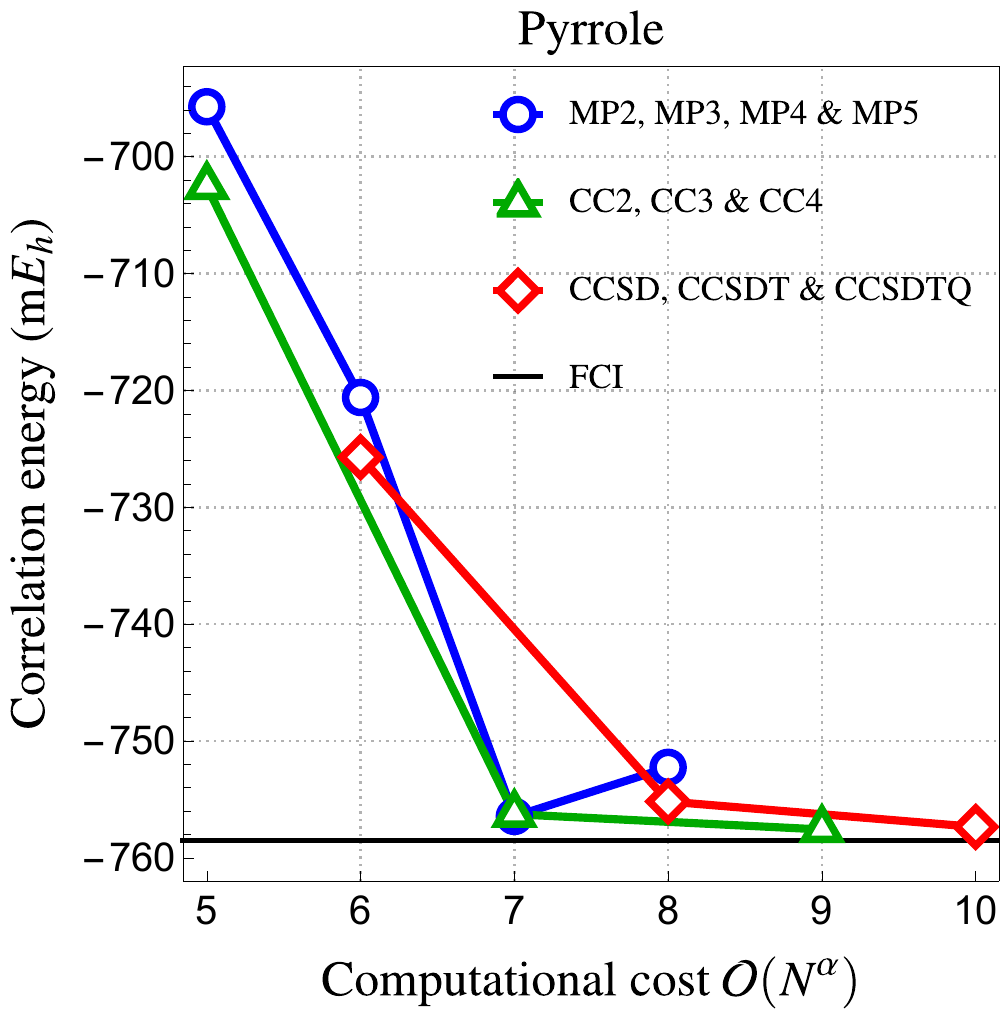}
	\\
	\includegraphics[width=0.24\textwidth]{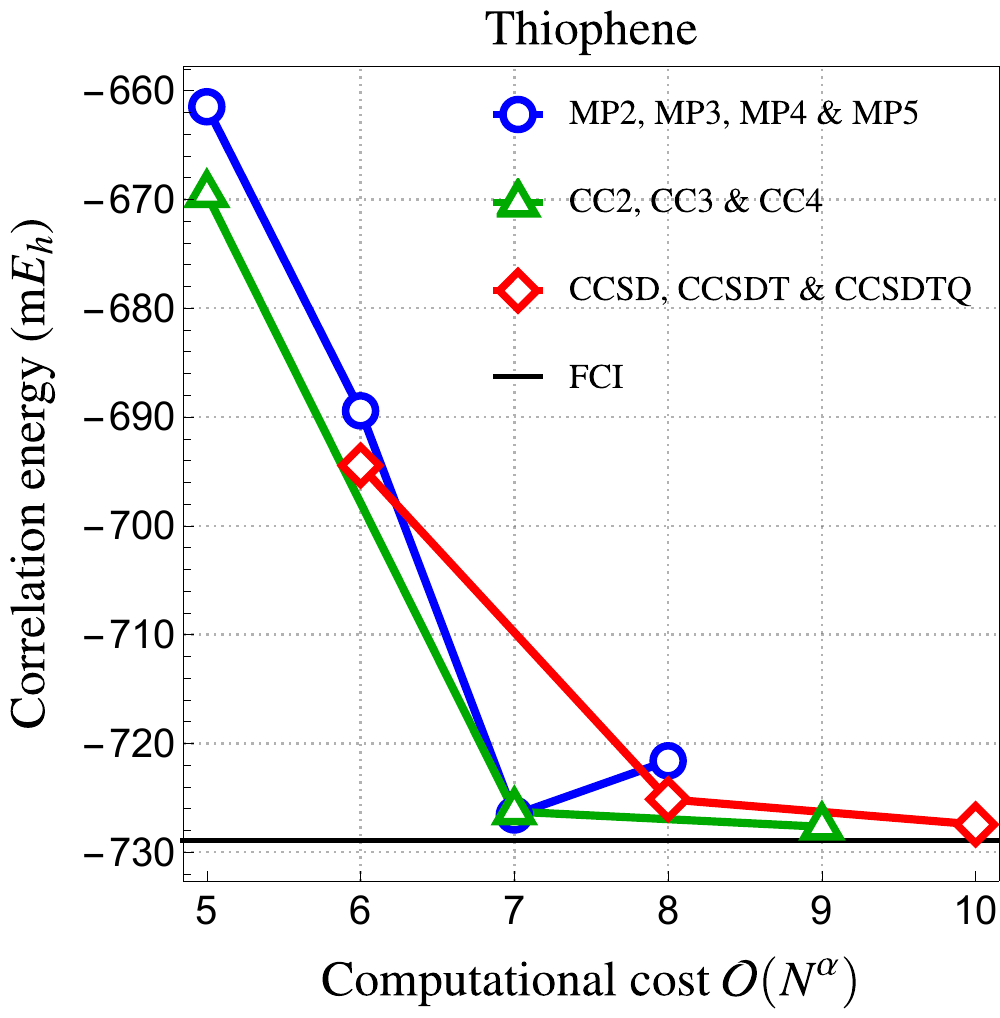}
	\includegraphics[width=0.24\textwidth]{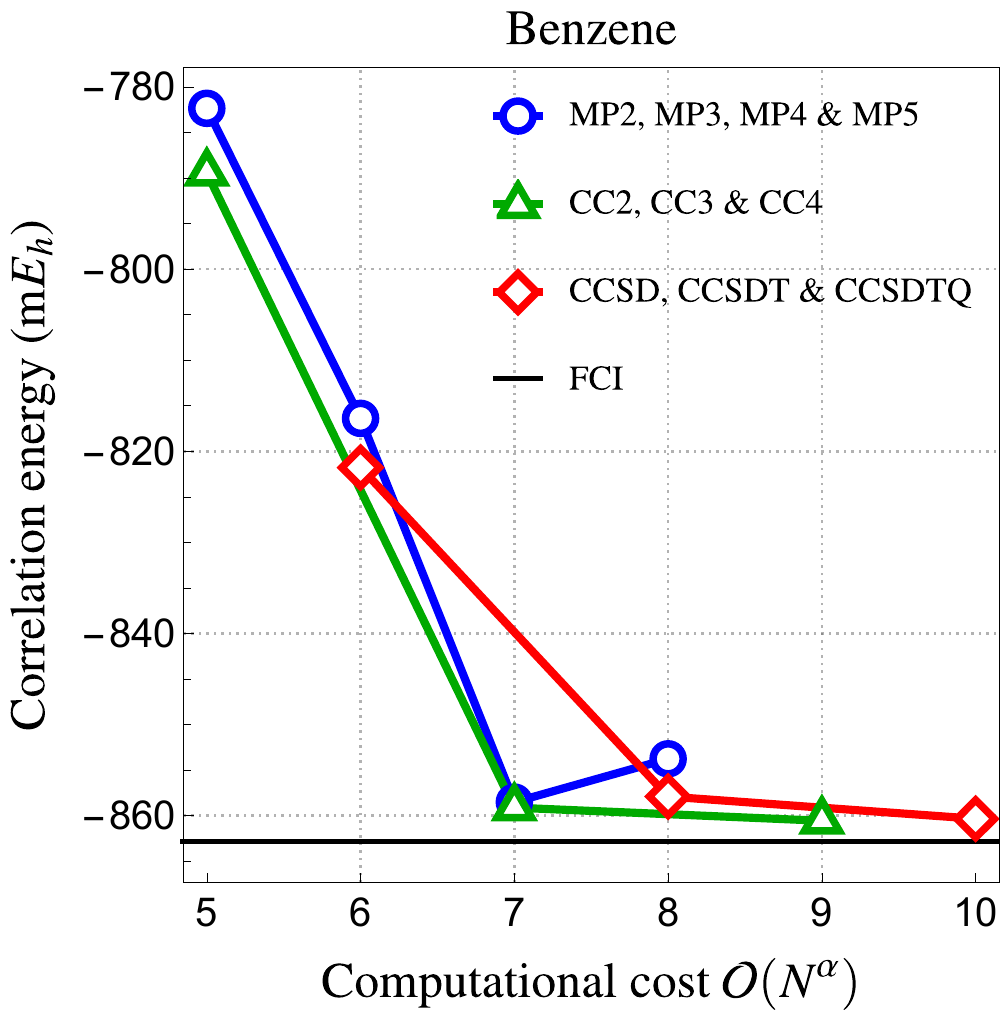}
	\includegraphics[width=0.24\textwidth]{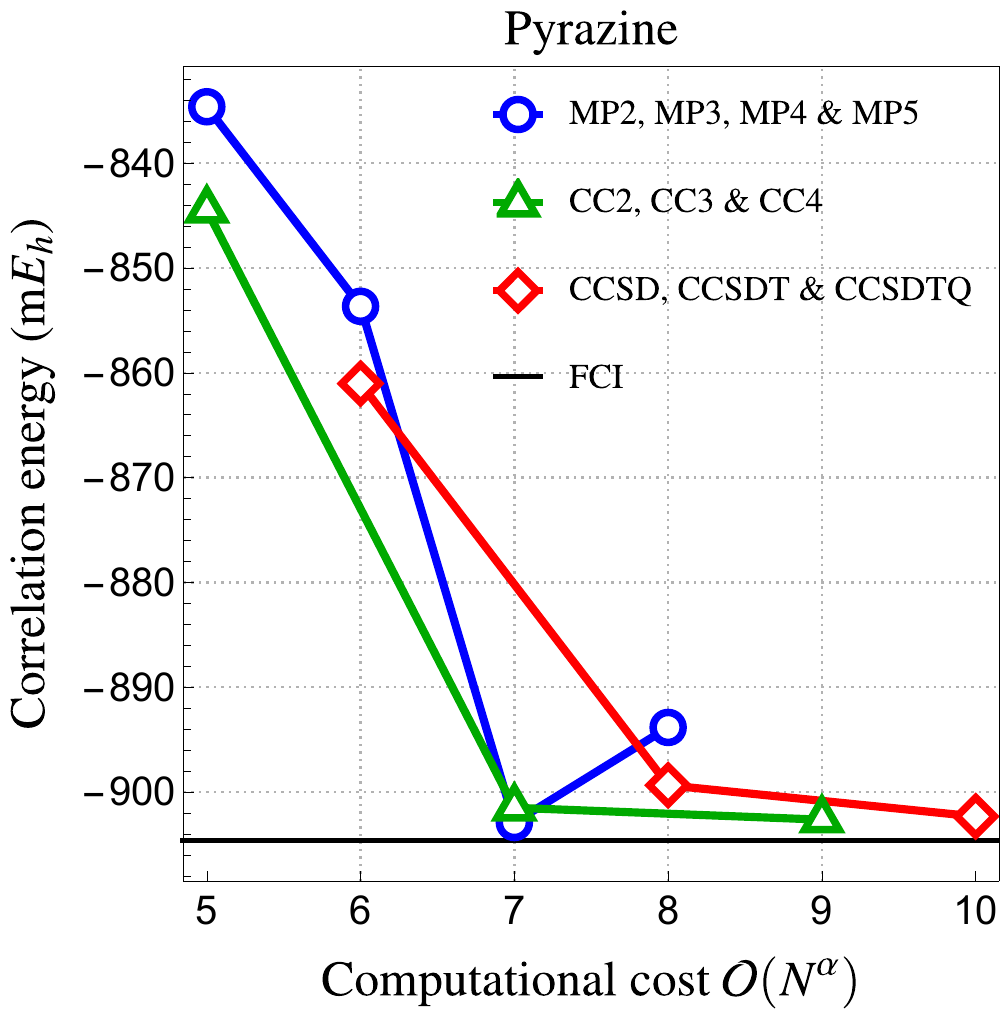}
	\includegraphics[width=0.24\textwidth]{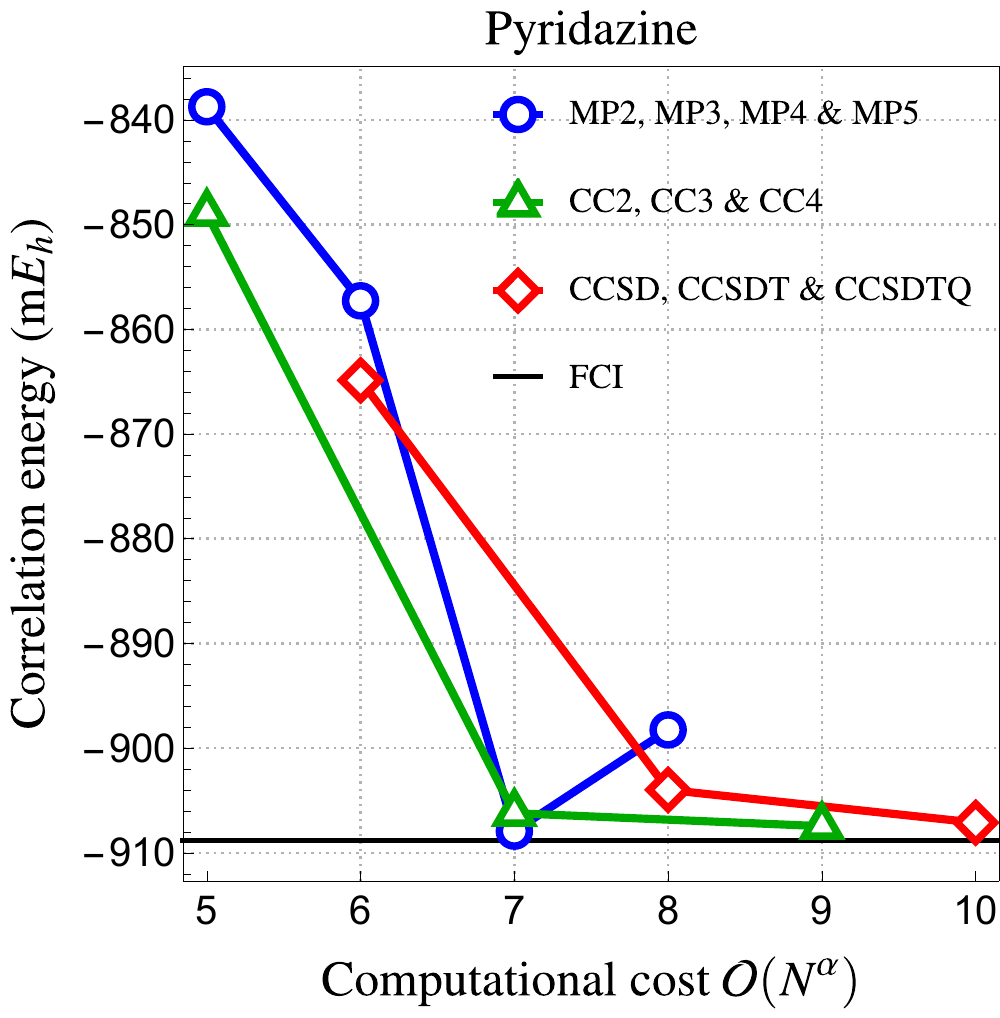}
	\\
	\includegraphics[width=0.24\textwidth]{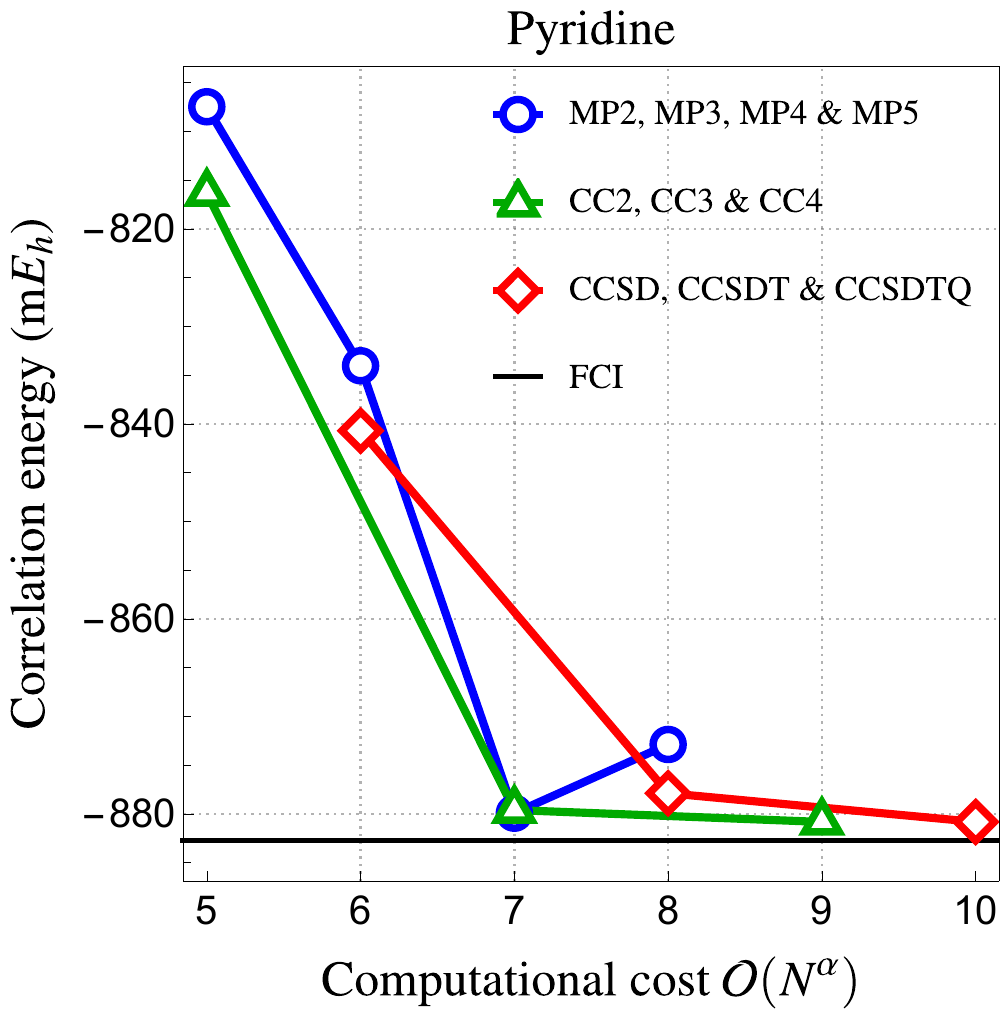}
	\includegraphics[width=0.24\textwidth]{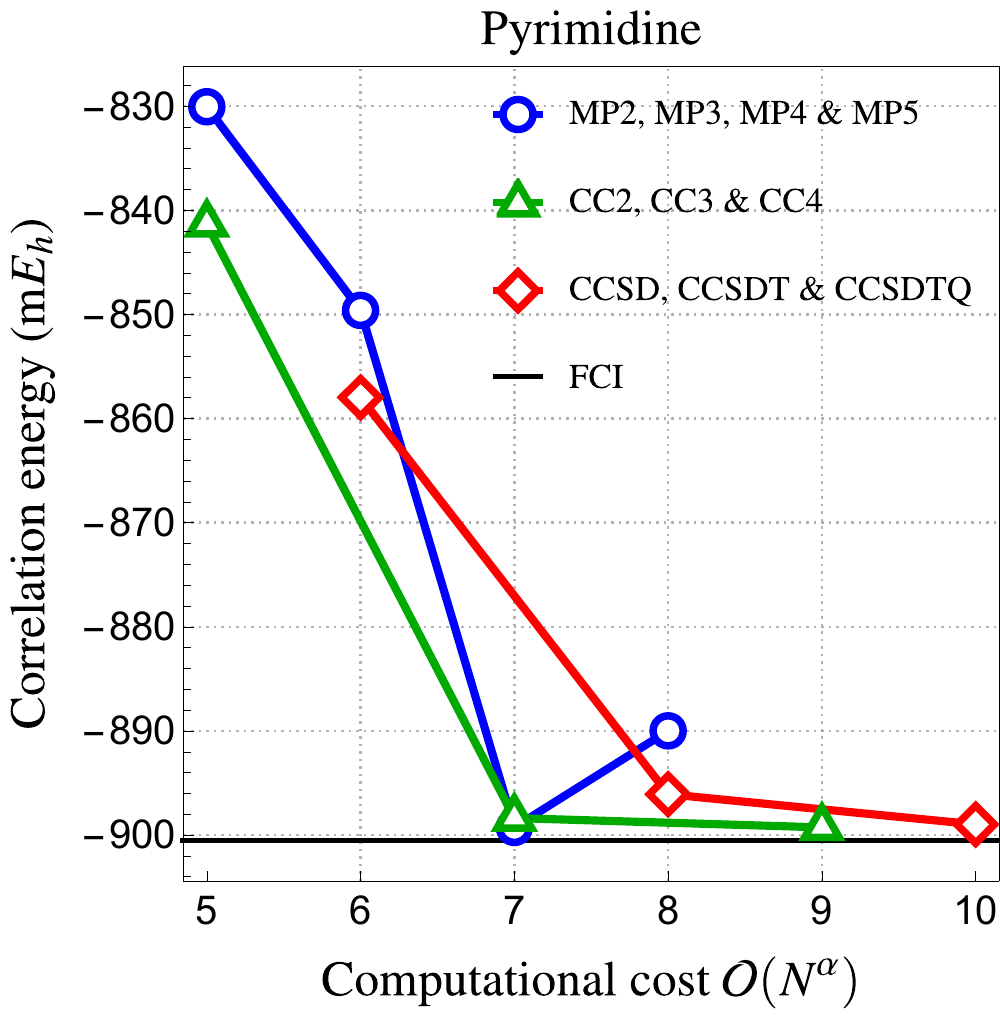}
	\includegraphics[width=0.24\textwidth]{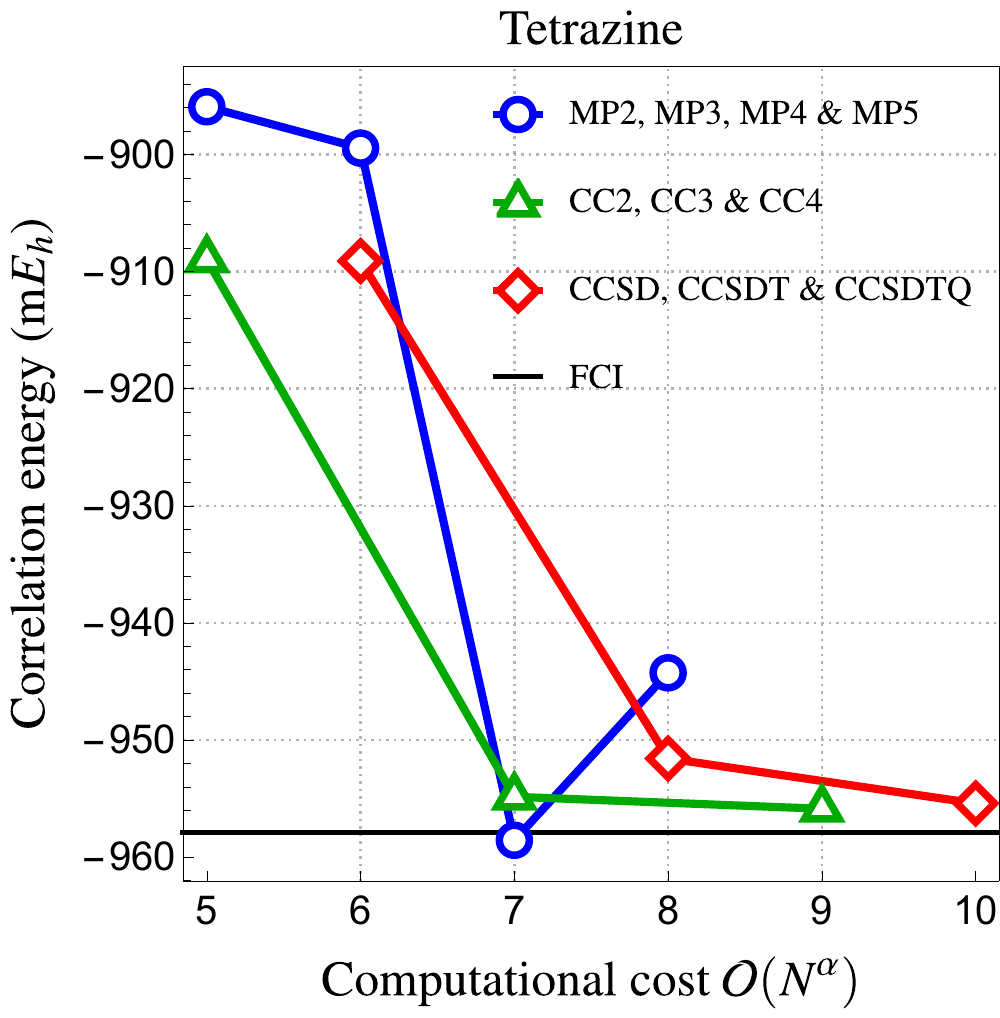}
	\includegraphics[width=0.24\textwidth]{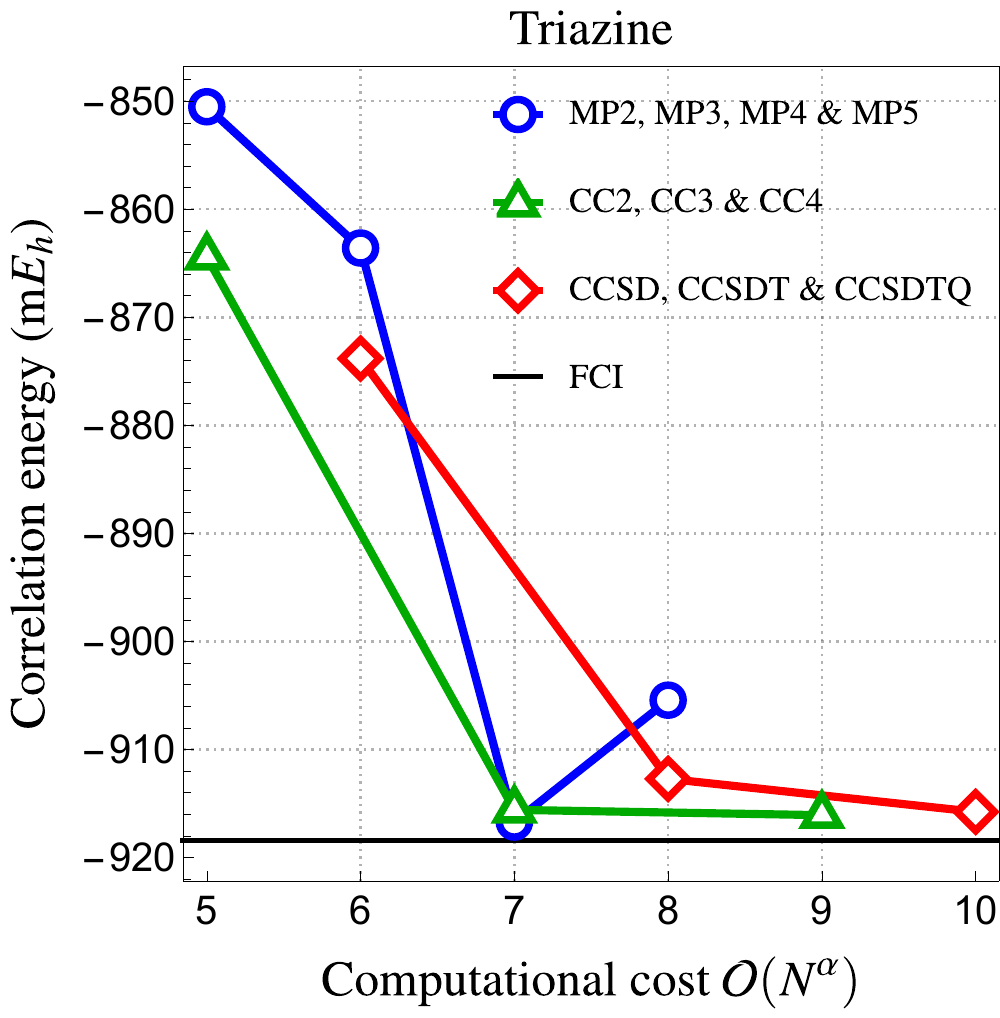}
	\caption{Convergence of the correlation energy (in \si{\milli\hartree}) computed in the cc-pVDZ basis as a function of the formal computational scaling for the twelve cyclic molecules represented in Fig.~\ref{fig:mol}.
	Three series of methods are considered: i) MP2, MP3, MP4, and MP5 (blue), ii) CC2, CC3, and CC4 (green), and iii) CCSD, CCSDT, CCSDTQ (red).
	The FCI estimate of the correlation energy is represented as a black line.
	\label{fig:MPCC}}
\end{figure*}

\begin{squeezetable}
\begin{table}
	\caption{
	Mean absolute error (MAE), mean signed error (MSE), and minimum (Min) and maximum (Max) absolute errors (in \si{\milli\hartree}) with respect to the FCI correlation energy computed in the cc-pVDZ basis for various methods. 
	The formal computational scaling of each method is also reported.
	\label{tab:stats}}
	\begin{ruledtabular}
	\begin{tabular}{lcdddd}
	Method	&	Scaling			&	\tabc{MAE}	&	\tabc{MSE}	&	\tabc{Max}	&	\tabc{Min}	\\
	\hline
	MP2		&	$\order{N^5}$	&	68.4	&	68.4	&	80.6	&	57.8	\\	
	MP3		&	$\order{N^6}$	&	46.5	&	46.5	&	58.4	&	37.9	\\
	MP4		&	$\order{N^7}$	&	2.1		&	2.0		&	4.7		&	0.7		\\
	MP5		&	$\order{N^8}$	&	9.4		&	9.4		&	13.6	&	5.8		\\
	\hline
	CC2		&	$\order{N^5}$	&	58.9	&	58.9	&	73.5	&	48.9	\\
	CC3		&	$\order{N^7}$	&	2.7		&	2.7		&	3.8		&	2.1		\\
	CC4		&	$\order{N^9}$	&	1.5		&	1.5		&	2.3		&	0.8		\\
	\hline
	CCSD	&	$\order{N^6}$	&	39.4	&	39.4	&	48.8	&	32.0	\\
	CCSDT	&	$\order{N^8}$	&	4.5		&	4.5		&	6.3		&	3.0		\\
	CCSDTQ	&	$\order{N^{10}}$&	1.8		&	1.8		&	2.6		&	1.0		\\
	\hline
	CCSD(T)		&	$\order{N^7}$	&	4.5		&	4.5		&	5.7		&	3.6		\\
	CR-CC(2,3)	&	$\order{N^7}$	&	5.0		&	5.0		&	6.6		&	3.6		\\
	\end{tabular}
	\end{ruledtabular}
\end{table}
\end{squeezetable}

\section{Computational details}
\label{sec:compdet}
The geometries of the twelve systems considered in the present study were all obtained at the CC3/aug-cc-pVTZ level of theory and were extracted from a previous study. \cite{Loos_2020b}
Note that, for the sake of consistency, the geometry of benzene considered here is different from the one of Ref.~\onlinecite{Loos_2020e} which was obtained at a lower level of theory [MP2/6-31G(d)]. \cite{Schreiber_2008}
The MP2, MP3, MP4, CC2, CC3, CC4, CCSD, CCSDT, and CCSDTQ calculations were performed with CFOUR, \cite{Matthews_2020} the CR-CC(2,3) calculations were made with GAMESS 2014R1, \cite{gamess} and MP5 and CCSD(T) calculations were computed with GAUSSIAN 09. \cite{g09} 
The CIPSI calculations were performed with QUANTUM PACKAGE. \cite{Garniron_2019}
In the current implementation, the selection step and the PT2 correction are computed simultaneously via a hybrid semistochastic algorithm.\cite{Garniron_2017,Garniron_2019} 
Here, we employ the renormalized version of the PT2 correction which was recently implemented and tested for a more efficient extrapolation to the FCI limit thanks to a partial resummation of the higher orders of perturbation. \cite{Garniron_2019}
We refer the interested reader to Ref.~\onlinecite{Garniron_2019} for further details.
For all these calculations, Dunning's correlation-consistent double-$\zeta$ basis (cc-pVDZ) has been employed.

Although the FCI energy has the enjoyable property of being independent of the set of one-electron orbitals used to construct the many-electron Slater determinants, as a truncated CI method, the convergence properties of CIPSI strongly dependent on this orbital choice.
In the present study, we investigate, in particular, the convergence behavior of the CIPSI energy for two sets of orbitals: natural orbitals (NOs) and optimized orbitals (OOs).
Following our usual procedure, \cite{Scemama_2018,Scemama_2018b,Scemama_2019,Loos_2018a,Loos_2019,Loos_2020a,Loos_2020b,Loos_2020c,Loos_2020e} we perform first a preliminary SCI calculation using HF orbitals in order to generate a SCI wave function with at least $10^7$ determinants.
Natural orbitals are computed based on this wave function and they are used to perform a new CIPSI run up to $8 \times 10^7$ determinants.
Successive orbital optimizations are then performed, which consist in minimizing the variational CIPSI energy at each macroiteration up to approximately $2 \times 10^5$ determinants.
When convergence is achieved in terms of orbital optimization, as our production run, we perform a new CIPSI calculation from scratch using this set of optimized orbitals to $8 \times 10^7$ determinants.
Using optimized orbitals has the undeniable advantage to produce, for a given variational energy, more compact CI expansions (see Sec.~\ref{sec:res}).
For the benzene molecule, we have also explored the use of localized orbitals (LOs) which are produced with the Boys-Foster localization procedure \cite{Boys_1960} that we apply to the natural orbitals in several orbital windows in order to preserve a strict $\sigma$-$\pi$ separation in the planar systems considered here. \cite{Loos_2020e}
Because they take advantage of the local character of electron correlation, localized orbitals have been shown to provide faster convergence towards the FCI limit compared to natural orbitals. \cite{Angeli_2003,Angeli_2009,BenAmor_2011,Suaud_2017,Chien_2018,Eriksen_2020,Loos_2020e}
As we shall see below, employing optimized orbitals has the advantage to produce an even smoother and faster convergence of the SCI energy toward the FCI limit.
Note that both localized and optimized orbitals do break the spatial symmetry.
Unlike excited-state calculations where it is important to enforce that the wave functions are eigenfunctions of the $\Hat{S}^2$ spin operator, \cite{Chilkuri_2021} the present wave functions do not fulfill this property as we aim for the lowest possible energy of a closed-shell singlet state.
We have found that $\expval*{\Hat{S}^2}$ is, nonetheless, very close to zero ($\sim 10^{-3}$) for each system.

The present CIPSI calculations have been performed on the AMD partition of GENCI's Irene supercomputer. 
Each Irene's AMD node is a dual-socket AMD Rome (EPYC) CPU at 2.60 GHz with 256GiB of RAM, with a total of 64 physical cores per socket. 
These nodes are connected via Infiniband HDR100. 
In total, the present calculations have required around 3~million core hours.

All the data (geometries, energies, etc) and supplementary material associated with the present manuscript are openly available in Zenodo at \url{http://doi.org/10.5281/zenodo.5150663}.

\section{Results and discussion}
\label{sec:res}

\subsection{CIPSI estimates}
\label{sec:cipsi_res}

We first study the convergence of the CIPSI energy as a function of the number of determinants.
Our motivation here is to generate FCI-quality reference correlation energies for the twelve cyclic molecules represented in Fig.~\ref{fig:mol} in order to benchmark the performances of various mainstream MP and CC methods (see Sec.~\ref{sec:mpcc_res}).
For the natural and optimized orbital sets, we report, in Fig.~\ref{fig:vsNdet}, the evolution of the variational correlation energy $\Delta \Evar = \Evar - \EHF$ (where $\EHF$ is the HF energy) and its perturbatively corrected value $\Delta \Evar + \EPT$ with respect to the number of determinants $\Ndet$ for each cyclic molecule.
As compared to natural orbitals (solid red lines), one can see that, for a given number of determinants, the use of optimized orbitals greatly lowers $\Delta \Evar$ (solid blue lines).
Adding the perturbative correction $\EPT$ yields very similar curves for both sets of orbitals (dashed lines).
This indicates that, for a given number of determinants, $\EPT$ (which, we recall, provides a qualitative idea to the distance to the FCI limit) is much smaller for optimized orbitals than for natural orbitals.
This is further evidenced in Fig.~\ref{fig:vsEPT2} where we show the behavior of $\Delta \Evar$ as a function of $\EPT$ for both sets of orbitals.
From Fig.~\ref{fig:vsEPT2}, \alert{it is clear one produces smaller $\EPT$ values when optimized orbitals are selected, hence facilitating the extrapolation procedure to the FCI limit (see below).}
The five-point weighted linear fit using the five largest variational wave functions are also represented (dashed black lines), while the FCI estimate of the correlation energy (solid black line) is reported for reference in Figs.~\ref{fig:vsNdet} and \ref{fig:vsEPT2}.

Figure \ref{fig:BenzenevsNdet} compares the convergence of $\Delta \Evar$ for natural, localized, and optimized orbitals for benzene.
As mentioned in Sec.~\ref{sec:compdet}, although both the localized and optimized orbitals break the spatial symmetry to take advantage of the local nature of electron correlation, the latter set further improves on the use of former set.
More quantitatively, optimized orbitals produce the same variational energy as localized orbitals with, roughly, a ten-fold reduction in the number of determinants.
A similar improvement is observed going from natural to localized orbitals.
According to these observations, all our FCI correlation energy estimates have been produced from the set of optimized orbitals.

To this end, we have extrapolated the orbital-optimized variational CIPSI correlation energies to $\EPT = 0$ via a weighted five-point linear fit using the five largest variational wave functions (see Fig.~\ref{fig:vsEPT2}).
The fitting weights have been taken as the inverse square of the perturbative corrections.
Our final FCI correlation energy estimates are reported in Tables \ref{tab:Tab5-VDZ} and \ref{tab:Tab6-VDZ} for the five- and six-membered rings, respectively, alongside their corresponding fitting error.
The stability of these estimates are illustrated by the results gathered in Table \ref{tab:fit}, where we list the extrapolated correlation energies $\Delta \Eextrap$ and their associated fitting errors obtained via weighted linear fits varying the number of fitting points from $3$ to $7$.
\alert{The extrapolation distance $\Delta \Edist$ defined as the difference between the final computed energy $\Delta \Efinal$ and $\Delta \Eextrap$ is also reported.}
Although we cannot provide a mathematically rigorous error bar, the data provided by Table \ref{tab:fit} show that the extrapolation procedure is robust and that our FCI estimates \alert{carry an error of the order of one millihartree}.
Logically, the FCI estimates for the five-membered rings seem slightly more accurate than for the (larger) six-membered rings.
It is pleasing to see that, although different geometries are considered, our present estimate of the frozen-core correlation energy of the benzene molecule in the cc-pVDZ basis (\SI{-862.9}{\milli\hartree}) is very close to the one reported in Ref.~\onlinecite{Loos_2020e} (\SI{-863.4}{\milli\hartree}).

Table \ref{tab:fit} does report extrapolated correlation energies and fitting errors for both natural and optimized orbitals.
Again, the superiority of the latter set is clear as both the variation in extrapolated values and the fitting error are much larger with the natural set.
\alert{Moreover, the extrapolation distance $\Delta \Edist$ is systematically decreases by several \si{\milli\hartree}.}
Taking cyclopentadiene as an example, the extrapolated values vary by almost \SI{1}{\milli\hartree} with natural orbitals and less than \SI{0.1}{\milli\hartree} with the optimized set. 
The fitting errors follow the same trend.

\subsection{Benchmark of CC and MP methods}
\label{sec:mpcc_res}

Using the CIPSI estimates of the FCI correlation energy produced in Sec.~\ref{sec:cipsi_res}, we now study the performance and convergence properties of three series of methods: i) MP2, MP3, MP4, and MP5, ii) CC2, CC3, and CC4, and iii) CCSD, CCSDT, and CCSDTQ.
Additionally, we also report CCSD(T) and CR-CC(2,3) correlation energies.
The raw data are reported in Tables \ref{tab:Tab5-VDZ} and \ref{tab:Tab6-VDZ} for the five- and six-membered rings, respectively. 
In Fig.~\ref{fig:MPCC}, we show, for each molecule, the convergence of the correlation energy for each series of methods as a function of the formal computational scaling of the corresponding method.
Statistical quantities [mean absolute error (MAE), mean signed error (MSE), minimum (Min) and maximum (Max) absolute errors with respect to the FCI reference values] are also reported in Table \ref{tab:stats} for each method as well as their formal computational scaling. 

First, we investigate the ``complete'' and well-established series of methods CCSD, CCSDT, and CCSDTQ.
Unfortunately, CC with singles, doubles, triples, quadruples, and pentuples (CCSDTQP) calculations are out of reach here. \cite{Hirata_2000,Kallay_2001}
As expected for the present set of weakly correlated systems, going from CCSD to CCSDTQ, one systematically and quickly improves the correlation energies with respective MAEs of $39.4$, $4.5$, \SI{1.8}{\milli\hartree} for CCSD, CCSDT, and CCSDTQ.
As usually observed, CCSD(T) (MAE of \SI{4.5}{\milli\hartree}) provides similar correlation energies than the more expensive CCSDT method by computing perturbatively (instead of iteratively) the triple excitations, while CCSD(T) and CR-CC(2,3) performs equally well.

Second, we investigate the approximate CC series of methods CC2, CC3, and CC4.
As observed in our recent study on excitation energies, \cite{Loos_2021} CC4, which returns a MAE of \SI{1.5}{\milli\hartree}, is an outstanding approximation to its CCSDTQ parent (MAE of \SI{1.8}{\milli\hartree}) and is, in the present case, even slightly more accurate in terms of mean errors as well as maximum and minimum absolute errors.
Moreover, we observe that CC3 provides very accurate correlation energies with a MAE of \SI{2.7}{\milli\hartree}, showing that this iterative method is particularly effective for ground-state energetics and outperforms both the perturbative CCSD(T) and iterative CCSDT models.
It is important to mention that even if the two families of CC methods studied here are known to be non-variational (see Sec.~\ref{sec:intro}), for the present set of weakly-correlated molecular systems, they never produce a lower energy than the FCI estimate as illustrated by the systematic equality between MAEs and MSEs.

Third, let us look into the MP series which is known, as mentioned in Sec.~\ref{sec:intro}, to potentially exhibit ``surprising'' behaviors depending on the type of correlation at play.\cite{Laidig_1985,Knowles_1985,Handy_1985,Gill_1986,Laidig_1987,Nobes_1987,Gill_1988,Gill_1988a,Lepetit_1988,Malrieu_2003}
(See Ref.~\onlinecite{Marie_2021a} for a detailed discussion).
For each system, the MP series decreases monotonically up to MP4 but raises quite significantly when one takes into account the fifth-order correction.
We note that the MP4 correlation energy is always quite accurate (MAE of \SI{2.1}{\milli\hartree}) and is only a few millihartree higher than the FCI value (except in the case of s-tetrazine where the MP4 number is very slightly below the reference value): MP5 (MAE of \SI{9.4}{\milli\hartree}) is thus systematically worse than MP4 for these weakly-correlated systems.
Importantly here, one notices that MP4 [which scales as $\order*{N^7}$] is systematically on par with the much more expensive $\order*{N^{10}}$ CCSDTQ method which exhibits a slightly smaller MAE of \SI{1.8}{\milli\hartree}.

\section{Conclusion}
\label{sec:ccl}
Using the SCI algorithm named \textit{Configuration Interaction using a Perturbative Selection made Iteratively} (CIPSI), we have produced FCI-quality frozen-core correlation energies for twelve cyclic molecules (see Fig.~\ref{fig:mol}) in the correlation-consistent double-$\zeta$ Dunning basis set (cc-pVDZ).
These estimates, which \alert{probably carry an error of the order of one millihartree}, have been obtained by extrapolating CIPSI energies to the FCI limit based on a set of orbitals obtained by minimizing the CIPSI variational energy.
Using energetically optimized orbitals, one can reduce the size of the variational space by one order of magnitude for the same variational energy as compared to natural orbitals.

Thanks to these reference FCI energies, we have then benchmarked three families of popular electronic structure methods: i) the MP perturbation series up to fifth-order (MP2, MP3, MP4, and MP5), ii) the approximate CC series CC2, CC3, and CC4, and iii) the ``complete'' CC series CCSD, CCSDT, and CCSDTQ.
With a $\order*{N^7}$ scaling, MP4 provides an interesting accuracy/cost ratio for this particular set of weakly correlated systems, while MP5 systematically worsen the perturbative estimates of the correlation energy.
In addition, CC3 (where the triples are computed iteratively) outperforms the perturbative-triples CCSD(T) method with the same $\order*{N^7}$ scaling, its completely renormalized version CR-CC(2,3), as well as its more expensive parent, CCSDT.
A similar trend is observed for the methods including quadruple excitations, where the $\order*{N^9}$ CC4 model has been shown to be slightly more accurate than CCSDTQ [which scales as $\order*{N^{10}}$], both methods providing correlation energies within \SI{2}{\milli\hartree} of the FCI limit.
\alert{These observations slightly alter the method ranking provided in Sec.~\ref{sec:intro}.
Of course, the present trends are only valid for this particular class of (weakly-correlated) molecules.
For example, the performance of CC3 might decline for larger systems.} 
Thus, it would be desirable to have a broader variety of systems in the future by including more challenging systems such as, for example, transition metal compounds.
Some work along this line is currently being performed.

As perspectives, we are currently investigating the performance of the present approach for excited states in order to expand the QUEST database of vertical excitation energies. \cite{Veril_2021} 
We hope to report on this in the near future.
The compression of the variational space brought by optimized orbitals could be also beneficial in the context of quantum Monte Carlo methods to generate compact, yet accurate multi-determinant trial wave functions. \cite{Dash_2018,Dash_2019,Scemama_2020,Dash_2021}

\section*{Supplementary Material}
Included in the supplementary material are the raw data for each figure, geometries, basis set files, orbitals obtained at various levels of theory, input and output files for each calculation, as well as a standalone \textsc{mathematica} notebook gathering modules for generating figures and statistics.

\begin{acknowledgements}
This work was performed using HPC resources from GENCI-TGCC (2021-gen1738), from the CCIPL computational center installed in Nantes, and from CALMIP (Toulouse) under allocation 2021-18005, and was also supported by the European Centre of Excellence in Exascale Computing TREX --- Targeting Real Chemical Accuracy at the Exascale. This project has received funding from the European Union's Horizon 2020 --- Research and Innovation program --- under grant agreement no.~952165.
This project has received funding from the European Research Council (ERC) under the European Union's Horizon 2020 research and innovation programme (Grant agreement No.~863481).
\end{acknowledgements}

\section*{Data availability statement}
The data that support the findings of this study are openly available in Zenodo at \url{http://doi.org/10.5281/zenodo.5150663}.

\bibliography{Ec}

\end{document}